\documentclass[preprint]{revtex4-1}
 
\usepackage{graphics}
\usepackage{graphicx}
\usepackage{pstricks}
\usepackage{amsbsy}
\usepackage{amsmath}

\usepackage[normalem]{ulem}
\definecolor{green4}{RGB}{0,139,0}

\renewcommand{\vec}{\mathbf}

\renewcommand{\tensor}{\mathbf}
\newcommand{\vecs}{\boldsymbol}

\newcommand{\p}{\partial}
\newcommand{\hardd}{\mathrm{d}}
\newcommand{\fluc}[1]{\delta #1}
\newcommand{\average}[1]{#1_{\text{av}}}
\newcommand{\w}{\widetilde}

\begin{document}

\title{Hydrodynamic relaxations in dissipative particle dynamics}

\author{J. S. Hansen} \email{jschmidt@ruc.dk}
\affiliation{
  ``Glass and Time'', IMFUFA, Department of Science and Environment,
  Roskilde University, Postbox 260, DK-4000 Roskilde, Denmark 
}
\author{Michael L. Greenfield} 
\affiliation{
 Department of Chemical Engineering, University of Rhode Island,
 Kingston, Rhode Island 02881, United States 
}
\author{Jeppe C. Dyre} 
\affiliation{
  ``Glass and Time'', IMFUFA , Department of Science and Environment,
  Roskilde University, Postbox 260, DK-4000 Roskilde, Denmark 
}

\begin{abstract}
  This paper studies the dynamics of relaxation phenomena in the
  standard dissipative particle dynamics (DPD) model [Groot and
    Warren, JCP, 107:4423 (1997)]. Using fluctuating hydrodynamics as
  the framework of the investigation, we focus on the collective
  transverse and longitudinal dynamics. It is shown that classical
  hydrodynamic theory predicts the transverse dynamics at relative low
  temperatures very well when compared to simulation data, however,
  the theory predictions are, on the same length scale, less accurate
  for higher temperatures. The agreement with hydrodynamics depends on
  the definition of the viscosity, and here we find that the
  transverse dynamics are independent of the dissipative and random
  shear force contributions to the stress.  For high temperatures, the
  spectrum for the longitudinal dynamics is dominated by the Brillouin
  peak for large length scales and the relaxation is therefore
  governed by sound wave propagation and is athermal. This contrasts
  the results at lower temperatures and small length scale, where the
  thermal process is clearly present in the spectra. The
  Landau-Placzek ratio is lower than the classical model Lennard-Jones
  liquid, especially at higher temperatures. The DPD model, at least
  qualitatively, re-captures the underlying hydrodynamical mechanisms,
  and quantitative agreement is excellent at intermediate temperatures
  for the transverse dynamics.
\end{abstract}

\maketitle

\section{Introduction}
The dissipative particle dynamics (DPD) method
\cite{hoogerbrugge:epl:1992,espanol:epl:1995} is widely used to
perform mesoscale computer simulations of, e.g., polymer solutions
\cite{kong:jcp:1997}, spinodal decomposition \cite{novik:pre:2000},
fluid flows in micro- and nanopores
\cite{boromand:cpc:2015,ranjith:mn:2015}, and cell membrane damage
\cite{groot:biophys:2001}, just to name a few examples. A standard DPD
simulation involves a set of point particles interacting by three
different forces: a conservative, a dissipative, and a random force in
such a manner that momentum is conserved. The DPD particle can be
thought of as a collection of molecules moving in a coherent fashion
\cite{espanol:simu:2002}. The forces are often tweaked to mimic
specific fluidic systems, e.g., the particles can be connected with
spring forces to simulate polymer solutions and melts; see also the
review by Moeendarbary \emph{et al.\ }
\cite{moeendarbary:jam:2009}. Importantly, the interparticle
conservative force is weak and usually without a strong repulsive
core, in fact, the conservative force is not necessary in order to
obtain hydrodynamic behavior \cite{marsh:pre:1997,ripoll:jcp:2001}.

In the DPD model by Groot and Warren \cite{groot:jcp:1997}, the
conservative force is linear with respect to the distance between the
two point masses. This model is simple and very appealing; however, it
yields an unrealistic equation of state which is quadratic in density
\cite{groot:jcp:1997}. Also, the dissipative force depends only on the
position and velocity differences of the two interacting particles and
neglects shear forces \cite{espanol:jcp:2017}. Nevertheless, the
parameter space for this model is quite large and the physical
interpretation of the parameters is not always straightforward. For
example, the particle density can be chosen as a \emph{free} parameter
for a given system, and from this choice the conservative force
parameter can be estimated using the compressibility
\cite{groot:jcp:1997}. Interestingly, this so-called adaptive
parameter approach leads to a decreasing viscosity for decreasing
temperature \cite{boromand:cpc:2015}, which characterizes a gas
\cite{mcquarrie:book:1976}. This gaseous behavior is also manifested
by a Schmidt number of order unity \cite{groot:jcp:1997}, where the
Schmidt number is defined as the ratio between the kinematic viscosity
and the diffusion coefficient. Bocquet and Charlaix
\cite{bocquet:csr:2010} conjectured that classical hydrodynamics is
valid for wavevectors $k$ fulfilling
$k<\sqrt{2\pi\rho/\eta_0 \tau_s}$, where $\rho$ is the density,
$\eta_0$ the shear viscosity and $\tau_s$ is the relaxation time given
by the shear stress relaxation \cite{hansen:langmuir:2015}. From this
criterion one can see that in the low density limit (low Schmidt
number) the classical hydrodynamic theory will break down even at
large length scales as the viscosity and relaxation time are only
functions of temperaure in this limit.

The hydrodynamic properties for the DPD technique have been
thouroughly investigated in the past, see for example
Refs. \onlinecite{marsh:pre:1997,ripoll:jcp:2001}. However, as the DPD
model is widely used by the simulation community
\cite{keaveny:jcp:2005,backer:jcp:2005,reddy:pf:2009,boromand:cpc:2015}
at low Schmidt number, we believe it is important to investigate the
properties of the model by Groot and Warren under conditions where the
Schmidt number varies from unity to higher values typically
characterizing liquids like the model Lennard-Jones liquid used in
classical molecular dynamics.

It has been noted by several authors
\cite{espanol:pre:1995,marsh:pre:1997}, that the energy is not
conserved in the standard DPD model, and that it cannot be applied to
study systems characterized by a sustained temperature gradient on the
macroscopic time scales. However, the model does feature fast energy
relaxations and, as also concluded by Marsh et
al. \cite{marsh:pre:1997}, it can indeed be applied to investigate
these relaxations. We wish to include this here as it will provide
valuable insight into the underlying mechanisms of the DPD method in
general.

We base our investigation on Onsager's regression hypothesis, which
states that the regression of microscopically induced fluctuations in
equilibrium follows the macroscopic laws of small non-equilibrium
disturbances \cite{onsager:physrev:1931}, i.e., thermally induced
perturbations relax according to hydrodynamics. Typically, these
(fast) relaxations do not refer to hydrodynamic quantities like
density and momentum directly, but instead to the decay of the
associated correlation functions \cite{kadanoff:annphys:1963}, as
predicted by hydrodynamic theory. We derive these correlation
functions from basic fluctuating hydrodynamics theory as this may not
be known to the reader; also, we present it in a slightly different
form (albeit equivalent) from that of standard texts
\cite{alley:pra:1983,hansen:book:2006,boon:book:1991}. To make the
study manageable, we focus on a limited part of the parameter space of
the standard DPD model.

\section{The hydrodynamic relaxation functions}
In general, one can write the balance equation for any hydrodynamic
quantity per unit mass $\phi=\phi(\vec{r},t)$ at position $\vec{r}$
and time $t$ as \cite{degroot:book:1984}
\begin{equation}
  \frac{\partial \rho \phi}{\partial t} = \sigma^\phi -
  \vecs{\nabla}\vec{J}^\phi - \vecs{\nabla}\cdot(\rho
  \phi\vec{u}) \,  ,
  \label{eq:balanceform}
\end{equation}
where $\vec{u}$ is the streaming velocity, $\sigma^\phi$ the
production term, and $\vec{J}^\phi$ the flux of $\phi$. In the case
$\sigma^\phi = 0$ the quantity is locally conserved. The hydrodynamic
quantities we study here are the mass density,
$\rho = \rho(\vec{r},t)$, the streaming velocity,
$\vec{u}=\vec{u}(\vec{r},t)$, and the excess kinetic energy per unit
mass, $e=e(\vec{r},t)$; the latter quantity is defined as the
difference between the local and average kinetic energy per unit mass,
$m e(\vec{r},t)= E_{\text{kin}}(\vec{r},t) - \frac{3}{2}k_BT$, where
$m$ is the particle mass. Based on the microscopic hydrodynamic
operator formalism \cite{hansen:langmuir:2015} one can derive the
following the balance equations on the form of
Eq. (\ref{eq:balanceform}) in the absence of any external driving
forces
\begin{subequations}
  \label{eq:balance}
  \begin{align}
    \frac{\p \rho}{\p t} &= - \vecs{\nabla}\cdot\tensor{J}^m  -
    \vecs{\nabla} \cdot (\rho \vec{u})   \label{eq:balance:mass}\\
    \frac{\p \rho \vec{u}}{\p t} &= - \vecs{\nabla}\cdot\tensor{P}  -
    \vecs{\nabla} \cdot (\rho \vec{u} \vec{u}) \\
    \frac{\p \rho e}{\p t} &= \sigma^e - \vecs{\nabla}\cdot\tensor{J}^e -
    \vecs{\nabla} \cdot (\rho e \vec{u}) \label{eq:balance_e}
  \end{align}  
\end{subequations}
where $\tensor{J}^m$ is the mass flux tensor due to density gradients,
$\tensor{P}$ is the pressure tensor, and $\tensor{J}^e$ the excess
kinetic energy flux tensor.  Importantly, the excess kinetic energy
per unit mass, $e(\vec{r},t)$, is not a conserved quantity; hence, a
production term $\sigma^e$ appears in Eq.\
(\ref{eq:balance_e}). Furthermore, for the mass balance equation,
Eq. (\ref{eq:balance:mass}), we have decomposed the mass flux into two
parts; one due to thermal motion, $\vec{J}^m$, and one due to the
fluid advective motion, $\rho\vec{u}$.

The three quantities can be written as the sum of the constant average
part and the fluctuating part, i.e., $\rho = \average{\rho} + \fluc{\rho}$,
$\vec{u}=\fluc{\vec{u}}=(\fluc{u}_x, \fluc{u}_y, \fluc{u}_z)$, and
$e=\fluc{e}$ since the averages of the streaming velocity and excess
kinetic energy are zero. To first order in the fluctuations we have
\begin{equation}
  \rho\vec{u} = (\average{\rho} + \fluc{\rho})\fluc{\vec{u}}
  \approx \average{\rho}\fluc{\vec{u}} \  \ \ \text{and} \ \
  \rho e \approx \average{\rho}\fluc{e} \, .
  \label{eq:approxfluc}
\end{equation}
Using the framework of fluctuating hydrodynamics
\cite{zarate:book:2006}, we now introduce the linear constitutive
relations with stochastic forcing
\begin{subequations}
  \label{eq:const}
  \begin{align}
    \tensor{J}^m &= -D \vecs{\nabla} \rho + \fluc{\tensor{J}^m} 
    \label{eq:notfick} \\
    \tensor{P}   &= \left(p_{eq} - \eta_v (\vecs{\nabla} \cdot
    \vec{u})\right)\tensor{I} - 2\eta_0 \stackrel{os}{(\vecs{\nabla}\vec{u})} +
    \fluc\tensor{P}  \label{eq:const_newton}\\ 
    \tensor{J}^e &= - \frac{\lambda}{c_V} \vecs{\nabla} e +
    \fluc{\tensor{J}^e}   \label{eq:const_fourier}
  \end{align}
\end{subequations}
where $D$ is the mass flux diffusivity coefficient, $p$ is the normal
pressure, $\eta_v$ and $\eta_0$ the bulk and shear viscosities,
$\lambda$ the heat conductivity, $c_V$ the specific heat per unit
mass at constant volume, and $\stackrel{os}{(\vecs{\nabla}\vec{u})}$ is
the trace-less symmetric part of the strain rate tensor.

Equations (\ref{eq:const_newton}) and (\ref{eq:const_fourier}) are
just the constitutive relation for a Newtonian fluid and Fourier's law
of conduction \cite{mcquarrie:book:1976} with added stochastic
forcing. However, as we cannot in general ignore cross-correlation
effects on small time and length scale, it is noted that $D$ is not
the self-diffusion coefficient \cite{maginn:jpc:1993}. Since the mass
density and excess kinetic energy are scalars, that is of the same
parity, both fluxes in Eqs. (\ref{eq:notfick}) and
(\ref{eq:const_fourier}) can depend on the gradients of $\rho$
and $e$ according to Courier's principle
\cite{degroot:book:1984}. Here we follow Alley and Alder
\cite{alley:pra:1983} and model the cross coupling through the
production term $\sigma_e$ and the pressure $p_{eq}$.

In equilibrium the stochastic forcing term has a zero average
\cite{zarate:book:2006} and is uncorrelated with the hydrodynamic
quantities, e.g., $\langle \fluc{\tensor{J}^m}(\vec{r},t)
\fluc{\vec{u}(\vec{r'},t')}\rangle = \tensor{0}$. Substituting
Eqs. (\ref{eq:approxfluc}) and (\ref{eq:const}) into Eq.\
(\ref{eq:balance}), we arrive at the stochastic dynamics. To first
order in the fluctuations this is
\begin{subequations}
  \label{eq:linns}
  \begin{align}
    \frac{\partial}{\partial t}  \fluc{\rho} &= D\nabla^2 \fluc{\rho} -
    \average{\rho} \vecs{\nabla}\cdot \fluc{\vec{u}} -
    \vecs{\nabla} \cdot {\fluc{\tensor{J}^m}} \\ 
    \average{\rho}\frac{\partial }{\partial t} \fluc\vec{u}&=
    -\vecs{\nabla}\fluc{p}_{eq} + (\eta_v +
    \eta_0/3)\vecs{\nabla}(\vecs{\nabla}\cdot\fluc\vec{u}) + \eta_0
    \nabla^2 \fluc\vec{u} - \nabla \cdot \fluc\tensor{P}    \\
    \average{\rho}\frac{\partial }{\partial t} \fluc{e}&= \sigma_e +
    \frac{\lambda}{c_V} \nabla^2\fluc{e} - \vecs{\nabla} \cdot
         {\fluc{\tensor{J}^e}} 
  \end{align}
\end{subequations}
since the advective terms are of second order. More advanced
stochastic descriptions have been developed in order to, for example,
include elastic properties of the fluid
\cite{tabak:jam:2015,wang:jsc:2016}. For local thermodynamic
equilibrium, the pressure fluctuations can be written as
\cite{hansen:book:2006}
\begin{equation}
  \fluc{p} = \left( \frac{\partial p}{\partial \rho} \right)_T
  \fluc{\rho} + \left( \frac{\partial p}{\partial T} \right)_\rho
  \fluc{T} = \frac{1}{\average{\rho}\chi_T} \fluc{\rho} +
  \frac{\beta_V}{c_V} \fluc{e} \, ,\label{eq:flucp}
\end{equation}
where $\chi_T = -1/V\left(\partial V/\partial p\right)_T$ is the
isothermal compressibility,
$\beta_\text{V} = \left(\partial p/ \partial T\right)_\rho$ is the
thermal pressure coefficient, and $\fluc{e}=c_V \fluc{T}$. The
production term for the excess kinetic energy is given by Alley and
Alder \cite{alley:pra:1983}
\begin{equation}
  \sigma_e = \frac{T\beta_\text{V}}{\average{\rho}} \frac{\partial
    \fluc{\rho}}{\partial t} =  \frac{T\beta_\text{V}}{\average{\rho}}
  \left( D \nabla^2 \fluc{\rho} - \average{\rho} \vecs{\nabla} \cdot
    \fluc{\vec{u}} - \vecs{\nabla} \cdot
    {\fluc{\tensor{J}^m}}\right)   \label{eq:prod_e} \, . 
\end{equation}
Defining the Fourier transform as
\begin{equation}
  \w{f}(\vec{k},t)=\iiint_{-\infty}^\infty f(\vec{r},t) \,
  e^{-i\vec{k}\cdot\vec{r}} \, \hardd \vec{r} 
\end{equation}
and then substituting Eqs. (\ref{eq:flucp}) and (\ref{eq:prod_e}) into
Eq.\ (\ref{eq:linns}) gives, in Fourier space,  
\begin{subequations}
  \label{eq:ddd}
  \begin{align}
    \frac{\partial}{\partial t}  \w{\fluc{\rho}} &= -Dk^2
    \w{\fluc{\rho}} - i\average{\rho}  \vec{k} \cdot
    \w{\fluc{\vec{u}}} - i\vec{k} \cdot \w{\fluc{\vec{J}}}^m\\
    \average{\rho}\frac{\partial }{\partial t} \w{\fluc\vec{u}}&=
    -\frac{i\vec{k}}{\average{\rho}\chi_T}\w{\fluc{\rho}} - (\eta_v +
    \eta_0/3)\vec{k}(\vec{k}\cdot \w{\fluc\vec{u}}) - \eta_0
    k^2 \w{\fluc\vec{u}} - \frac{i \beta_V\vec{k}}{c_V}\w{\fluc{e}} -
    i\vec{k} \cdot \w{\fluc\tensor{P}} \\ 
    \average{\rho}\frac{\partial}{\partial t} \w{\fluc{e}} &=
    -\frac{T\beta_VDk^2}{\average{\rho}}\w{\fluc{\rho}} -
    iT\beta_V\vec{k}\cdot\fluc{\vecs{u}} - 
    \frac{\lambda k^2}{c_V} \w{\fluc{e}} - i \vec{k} \cdot \left(
    \w{\fluc{\tensor{J}}}^e + \w{\fluc{\tensor{J}}}^m \right)
  \end{align}
\end{subequations}
If one makes a particularly simple choice for the wavevector, then
the dynamics can be decomposed into transverse (normal) and
longitudinal (parallel) dynamics with respect to this wavevector.
For example, if we select $\vec{k}=(0, k, 0)$, then from
Eq.\ (\ref{eq:ddd}) the transverse dynamics is given by the streaming
velocity components $ \w{\fluc{u}}_x$ and $\w{\fluc{u}}_z$ via
\begin{subequations}
  \begin{align}
    \frac{\partial}{\partial t}  \w{\fluc{u}}_x&= -\nu_0 k^2
    \w{\fluc{u}}_x  - \frac{ik}{\average{\rho}}\w{\fluc{P}_{yx}} \label{eq:du} \\
    \frac{\partial }{\partial t} \w{\fluc{u}}_z&= -\nu_0 k^2
    \w{\fluc{u}}_z - \frac{ik}{\average{\rho}}\w{\fluc{P}_{yz}} 
     \label{eq:duz}
  \end{align}
\end{subequations}
where $\nu_0=\eta_0/\average{\rho}$ is the kinematic viscosity. We
will use both the dynamic viscosity, $\eta_0$, and kinematic
viscosity, $\nu_0$, whenever one is more convenient than the other. As
expected, Eqs. (\ref{eq:du}) and (\ref{eq:duz}) are identical with
respect to the dynamics and that the transverse dynamics are
independent of the energy and density fluctuations. The longitudinal
dynamics are given by
\begin{subequations}
  \label{eq:longdyn}
  \begin{align}
    \frac{\partial}{\partial t}  \w{\fluc{\rho}}&= -Dk^2\w{\fluc{\rho}} -
    i\average{\rho}k \w{\delta u}_y
    -ik\w{\fluc{J}}_y^m \label{eq:longdyn:rho}\\ 
    \frac{\partial}{\partial t}  \w{\fluc{u}}_y&=
    -\frac{ik}{\average{\rho}^2\chi_T} \w{\fluc{\rho}} -\nu_lk^2
    \w{\fluc{u}}_y - \frac{ik\beta_V}{c_V \average{\rho}}
    \w{\fluc{e}} -\frac{ik}{\average{\rho}}\w{\fluc{P}}_{yy}
    \label{eq:longdyn:vel} \\
    \frac{\partial }{\partial t} \w{\fluc{e}} &=
    -\frac{T\beta_VDk^2}{\average{\rho}^2} \w{\fluc{\rho}} -
    \frac{iT\beta_V k}{\average{\rho}}\w{\fluc{u}}_y -
    \kappa k^2 \w{\fluc{e}} -\frac{ik}{\average{\rho}}
    \left (\w{\fluc{J}}_y^e+\w{\fluc{J}}_y^m \right)  \label{eq:longdyn:energy}
  \end{align}
\end{subequations}
where $\nu_l=(\eta_v + 4\eta_0/3)/\average{\rho}$ is the longitudinal
kinematic viscosity and $\kappa=\lambda/(c_V \average{\rho})$.   

As mentioned above, one usually does not study the fluctuating
quantities directly, but rather the associated correlation
functions. To this end we define the equilibrium time-correlation
function between quantities $A$ and $B$ as
\begin{equation}
C_{AB}(\vec{k},t) = \frac{1}{V}\left\langle A(\vec{k}, t)
B(-\vec{k},0)\right\rangle   \, , \label{eq:cabdef}
\end{equation}
where $V$ is the system volume. Thus, multiplying Eqs. (\ref{eq:du})
with $\w{\fluc{u}}_x(-k,0)$ and taking the ensemble average over initial
conditions leads to
\begin{equation}
  \label{eq:aaa}
   \frac{\partial C_{uu}^\perp}{\partial t} = -\nu_0 k^2C_{uu}^\perp
\end{equation}
for the transverse relaxation.  Here $C_{uu}^\perp=\langle \w{\fluc{u}}_x(k,t)
\w{\fluc{u}}_x(-k,0)\rangle/V$ is the transverse velocity
autocorrelation function, and we have used that the stochastic
forcing term is uncorrelated with the fluctuating quantities. The
solution to Eq.\ (\ref{eq:aaa}) is
\begin{equation}
  \label{eq:trans:1}
  C_{uu}^{\perp}(k, t) = \frac{k_BT}{\average{\rho}} e^{-\nu_0 k^2 t} \, ,
\end{equation}
where the initial value $C_{uu}^\perp(k, 0)=k_BT/\average{\rho}$
is found from equipartition \cite{mcquarrie:book:1976}.

From Eq.\ (\ref{eq:longdyn}) one can form nine coupled correlation
functions for the longitudinal dynamics. For example, dynamic
equations for $C_{\rho\rho}, C_{\rho u}, C_{\rho e}$ are formed by
multiplying Eq.\ (\ref{eq:longdyn:rho}) with
$\fluc{\w{\rho}}(-\vec{k},0), \fluc{\w{u}}(-\vec{k},0)$, and
$\fluc{\w{e}}(-\vec{k},0)$, respectively, and taking the ensemble
average. In matrix notation, using the definition in
Eq. (\ref{eq:cabdef}) yields the following coupled linear differential
equation system
\begin{equation}
  \label{eq:CABeqns}
  \frac{\hardd }{\hardd t}
  \begin{bmatrix}
    C_{\rho\rho} & C_{\rho u} & C_{\rho e} \\
    C_{u\rho} & C_{u u} & C_{u e}        \\
    C_{e\rho} & C_{e u} & C_{e e} 
  \end{bmatrix}
  = -
  \begin{bmatrix}
    Dk^2 & i\average{\rho}k & 0 \\
    \frac{ik}{\average{\rho}^2\chi_T}  & \nu_lk^2 &
    \frac{ik\beta_V}{c_V \average{\rho}} \\
    \frac{T\beta_VDk^2}{\average{\rho}^2} &  \frac{iT\beta_V
      k}{\average{\rho}} &  \kappa k^2
  \end{bmatrix}
 \begin{bmatrix}
    C_{\rho\rho} & C_{\rho u} & C_{\rho e} \\
    C_{u\rho} & C_{u u} & C_{u e}        \\
    C_{e\rho} & C_{e u} & C_{e e} 
  \end{bmatrix} \, .
\end{equation}
The coefficient matrix is referred to as the hydrodynamic matrix
\cite{hansen:book:2006}. By performing the matrix multiplication in
Eq.\ (\ref{eq:CABeqns}) it is seen that the longitudinal dynamics
can be divided into three sets of co-dependent correlation functions,
for example, $\dot{C}_{\rho\rho} = A_1(C_{\rho\rho}, C_{u\rho})$,
$\dot{C}_{u\rho} = A_2(C_{\rho\rho}, C_{u\rho}, C_{e\rho})$, and
$\dot{C}_{e\rho} = A_3(C_{\rho\rho}, C_{u\rho}, C_{e\rho})$, where $A_1, A_2$ and
$A_3$ are linear functions represented by the hydrodynamic matrix. The
three sets are written as triplets
\begin{equation}
\{C_{\rho\rho}, C_{u\rho}, C_{e\rho}\}, \ 
\{C_{uu}, C_{\rho u},C_{e u}\},  \ \text{and} \ 
\{C_{ee}, C_{\rho e}, C_{ue}\} \, 
\end{equation}
and each set of coupled differential equations can be solved
from the hydrodynamic matrix. Up to second order in the wavevector,
the solution for any of the nine correlation functions has the form
\begin{equation}
  \label{eq:long}
  C_{AB}(\vec{k},t) = K_1e^{-D_Tk^2 t} +
  e^{-\Gamma k^2t} \left[ K_2 \cos(c_skt) + iK_3\sin(c_skt)\right]
\end{equation}
where   
\begin{equation}
  \label{eq:eigenvalues}
  D_T = \frac{\kappa}{\chi_T \average{\rho} c_s^2}
  \  \ \mathrm{and} \ \ \Gamma =  \frac{1}{2}\left[ \frac{\kappa}{\chi_T\average{\rho}c_s^2} +
    (D+\nu_l+\kappa) \right] 
\end{equation}
are the thermal diffusivity and sound attenuation, respectively, and
$c_s$ defined as
\begin{equation}
  \label{eq:cs_defn}
  c_s^2 = \frac{\beta_V^2\chi_TT - \average{\rho}c_V}{\chi_Tc_V\average{\rho}^2}
\end{equation}
is the adiabatic speed of sound. The three integrating factors $K_1,
K_2,$ and $K_3$ are found from the initial conditions and are, in
fact, not independent. Now, $C_{AB}$ is either a real or purely
imaginary valued function, which means that if $K_3 = 0$ then in
general $K_2 \neq 0$ and $K_1 \neq 0$\, while
if $K_3\neq 0$ then $K_2 = K_1 = 0$. In the case where $C_{AB}$ is
real, the normalized correlation function is written in
the form
\begin{equation}
  \label{eq:longnorm}
  C_{AB}^N(\vec{k},t) = K_{AB}e^{- D_T k^2 t} +
  (1-K_{AB})e^{-\Gamma k^2t}\cos(c_skt) \, . 
\end{equation}
Thus, the longitudinal dynamics are governed by three fundamental
processes with frequencies $D_Tk^2, \Gamma k^2$, and $c_sk$. From Eq.\
(\ref{eq:eigenvalues}), one sees that $D_T$ pertains to the thermal
processes and that the sound attenuation $\Gamma$ dampens the wave
propagation with speed $c_s$; the magnitude of this damping is
governed by all three diffusive processes, i.e., by $D, \nu_l$, and
$\kappa$. Equations (\ref{eq:trans:1}) and (\ref{eq:longnorm}) form
the framework for this hydrodynamics study.


\section{Simulation methodology}
\label{sect:model}
The standard DPD model by Groot and Warren is composed of a single
type of point particle. The particle position, $\vec{r}_i$, and
momentum, $\vec{p}_i$, follow Newton's equation of motion,
\begin{subequations}
  \begin{align}
    \frac{\hardd \vec{r}_i}{\hardd t} &= \frac{\vec{p}_i}{m} \\
    \frac{\hardd \vec{p}_i}{\hardd t} &= \vec{F}_i \, .
  \end{align}
\end{subequations}
The total force, $\vec{F}_i$, is composed of the conservative force,
$\vec{F}_i^C$, due to the interaction between the particles, a random
force, $\vec{F}_i^R,$ simulating the coarse graining of many degrees
of freedom, and a dissipative force, $\vec{F}_i^D$, removing the
viscous heating generated from the random force. Thus
$\vec{F}_i= \vec{F}_i^C + \vec{F}_i^R+ \vec{F}_i^D$. As it is common
practise, we use reduced units such that the characteristic mass and
length scales are set to unity. Also, temperature, $T$, is in units of
$k_B/\epsilon$, where $\epsilon$ is the characteristic energy
scale. In reduced units the conservative force is
\begin{equation}
  \vec{F}_{ij}^C = a_{ij}(1-r_{ij})\hat{\vec{r}}_{ij} \, ,
\end{equation}
where $a_{ij}$ is a parameter that quantifies the repulsion between
particles $i$ and $j$, $\vec{r}_{ij}$ is the vector of separation
$\vec{r}_i-\vec{r}_j$, $r_{ij}=|\vec{r}_{ij}|$, $\hat{\vec{r}}_{ij} =
\vec{r}_{ij}/r_{ij}$. Here we use $a_{ij}=25$ and the interactions are
ignored when $r_{ij}>1=r_c$.  Following Groot and Warren
\cite{groot:jcp:1997}, the random and dissipative forces are
\begin{equation}
  \vec{F}_{ij}^R = 
  \frac{\sigma w(r_{ij})\zeta_{ij}}{\sqrt{\Delta t}}
  \hat{\vec{r}}_{ij} \ \ \text{and} \ \ 
  \vec{F}_{ij}^D = -\frac{(\sigma w(r_{ij}))^2}{2mT}
  \left[\hat{\vec{r}}_{ij} \cdot
    (\vec{v}_i-\vec{v}_j)\right]\hat{\vec{r}}_{ij}  \ \ ,
\end{equation}
where $\sigma$ is the random force amplitude, $\zeta_{ij}$ is a
uniformly distributed random number with zero mean and unit variance,
$w(r_{ij})$ is a weighing function given by $w(r_{ij})=1-r_{ij}$,
$\vec{v}_i$ the velocity of particle $i$, and $\Delta t =0.02$ is the
time step used in the integrator. In all simulations the amplitude
$\sigma$ is set to 3.0.  The equations of motion are integrated
using the standard velocity Verlet algorithm by Groot
and Warren \cite{groot:jcp:1997}. The system size is 5000 particles at
density $\average{\rho}=3.0$, and temperatures (in reduced units) in
the range $0.1 \leq T \leq 1.0$ are simulated.

Espa\~{n}ol and Serrano \cite{espanol:pre:1999} studied the DPD model
in terms of dimensionless parameters, namely, friction,
$\Omega=\sigma^2r_c/(6v_Tk_BTm)$ where $v_T=\sqrt{k_BT/m}$, an overlap
parameter, $s=r_c \rho^{1/3}$, and system length scale,
$\mu=L_{\mathrm{box}}/r_c$. For relatively large friction and overlap
the particle dynamics are affected by the surrounding fluid, that is,
one would expect strong collective hydrodynamics. On the other hand,
for low friction and small overlap the dynamics are characterized by
single particle properties described by what Espa\~{n}ol and Serrano
call kinetic theory \cite{espanol:pre:1999}. In the simulations
carried out here, we only vary the temperature giving
$1.5 \leq \Omega \leq 14.7$, $s \approx 1.4$ and $\mu \approx 6.9$,
and we span both the kinetic (high $T$) and hydrodynamic regime (low
$T$).

During the simulations, all ten correlation functions are evaluated
from the microscopic definition of the hydrodynamic variables, which
to first order in fluctuations are
\begin{subequations}
  \begin{align}
    \w{\rho}(\vec{k},t) &= \sum_i m
                          e^{-i\vec{k}\cdot\vec{r}_i(t)}\\
    \w{\fluc{\vec{u}}}(\vec{k},t) &=
    \frac{1}{\average{\rho}}\sum_i m \vec{v}_i
                          e^{-i\vec{k}\cdot\vec{r}_i(t)}\\
    \w{\fluc{e}}(\vec{k},t) &=
    \frac{1}{\average{\rho}}\left[\sum_i \frac{1}{2}m v_i^2
    e^{-i\vec{k}\cdot\vec{r}_i(t)} - \frac{3}{2}k_BT\right]
  \end{align}
\end{subequations}
The viscosity at zero wavevector and frequency is also
evaluated. Recently, based on generic projection methods
\cite{espanol:ptrs:2002,ernst:epl:2006} Jung and Schmid
\cite{jung:jcp:2016} argued that the correct Green-Kubo integral is
\begin{equation}
  \eta_0^2 = \frac{V}{3k_BT}\left [ \frac{1}{2}
    \Delta t \sum_{\alpha\beta} \langle P_{\alpha\beta}^R(0)^2\rangle + \int_0^\infty
     \sum_{\alpha\beta} \left \langle
    (P_{\alpha\beta}^C(0)-P_{\alpha\beta}^D(0))(P_{\alpha\beta}^C(t)+P_{\alpha\beta}^D(t))
    \right \rangle \mathrm{d} t \right] \, , \label{eq:viscjs}
\end{equation}
where the double index $\alpha\beta$ runs over the $xy$, $xz$, and
$yz$ components of the pressure tensor; superscript 2 on $\eta$
distinguishes it from a viscosity defined by Groot and Warren
\cite{groot:jcp:1997} and used below.  $P_{\alpha\beta}^C$ are the
three off-diagonal elements of the Irving-Kirkwood pressure tensor
\cite{irving:jcp:1950}
\begin{equation}
  \label{eq:stressdef}
   V \tensor{P}^C(t) =
    \sum_i\frac{\vec{p}_i\vec{p}_i}{m_i} +
    \sum_i\sum_{j>i}\vec{r}_{ij}\vec{F}^C_{ij}\, ,
\end{equation}
and $P_{\alpha\beta}^D$ and $P_{\alpha\beta}^R$ are the dissipative and
random off-diagonal components of the tensors
\begin{equation}
   V \tensor{P}^D(t) = 
   \sum_i\sum_{j>i}\vec{r}_{ij}\vec{F}^D_{ij}
   \ \ \mathrm{and} \ \ 
  V \tensor{P}^R(t) = 
   \sum_i\sum_{j>i}\vec{r}_{ij}\vec{F}^R_{ij} \, .
\end{equation}
Other authors have evaluated the viscosity based on the Irving-Kirkwood
pressure only
\begin{equation}
  \eta_0^1 = \frac{V}{3k_BT}\int_0^\infty \left \langle
    P_{\alpha\beta}^C(0) P_{\alpha\beta}^C(t) \right \rangle
    \mathrm{d} t \, . \label{eq:viscik}
\end{equation}
We will compare the predictions from the hydrodynamic theory using
both definitions, Eqs. (\ref{eq:viscjs}) and (\ref{eq:viscik}). The
complex viscosity is calculated from the Irving-Kirkwood pressure
tensor, i.e., 
\begin{equation}
  \label{eq:etaomega}
  \eta^*(\omega) = \frac{V}{3k_BT}\int_0^\infty
  \sum_{\alpha\beta}\langle P^C_{\alpha\beta}(t)P^C_{\alpha\beta}(0)\rangle \,
  e^{-i\omega t} \, \hardd t \, .
\end{equation}
Finally, the self-diffusivity coefficient, $D_s$, is evaluated from
the Green-Kubo integral of the single particle velocity
autocorrelation function. We find that this leads to lower statistical
uncertainties compared to evaluating $D_s$ using the particle
mean-square displacements.

In a few cases, the dynamics of the DPD model is compared to a
liquid-phase Lennard-Jones system at the state-point
$(\rho, T)=(0.85,1.121)$ in units of $\sigma^3$ and
$k_B/\epsilon$. The Lennard-Jones particles interact through the
standard shifted 12-6 potential \cite{lennard-jones:prsla:1924} using
a cut-off distance at $r/\sigma=2.5$. The system size is $N=1000$, and
the equations of motion are integrated using the
leap-frog method \cite{frenkel:book:1996}. To control the temperature,
the Nose-Hoover thermostat \cite{nose:molphys:1984,hoover:pra:1985} is
applied. The dynamic properties are calculated as explained above.

\section{Results and Discussion}
\begin{figure}
  \includegraphics[scale=.25]{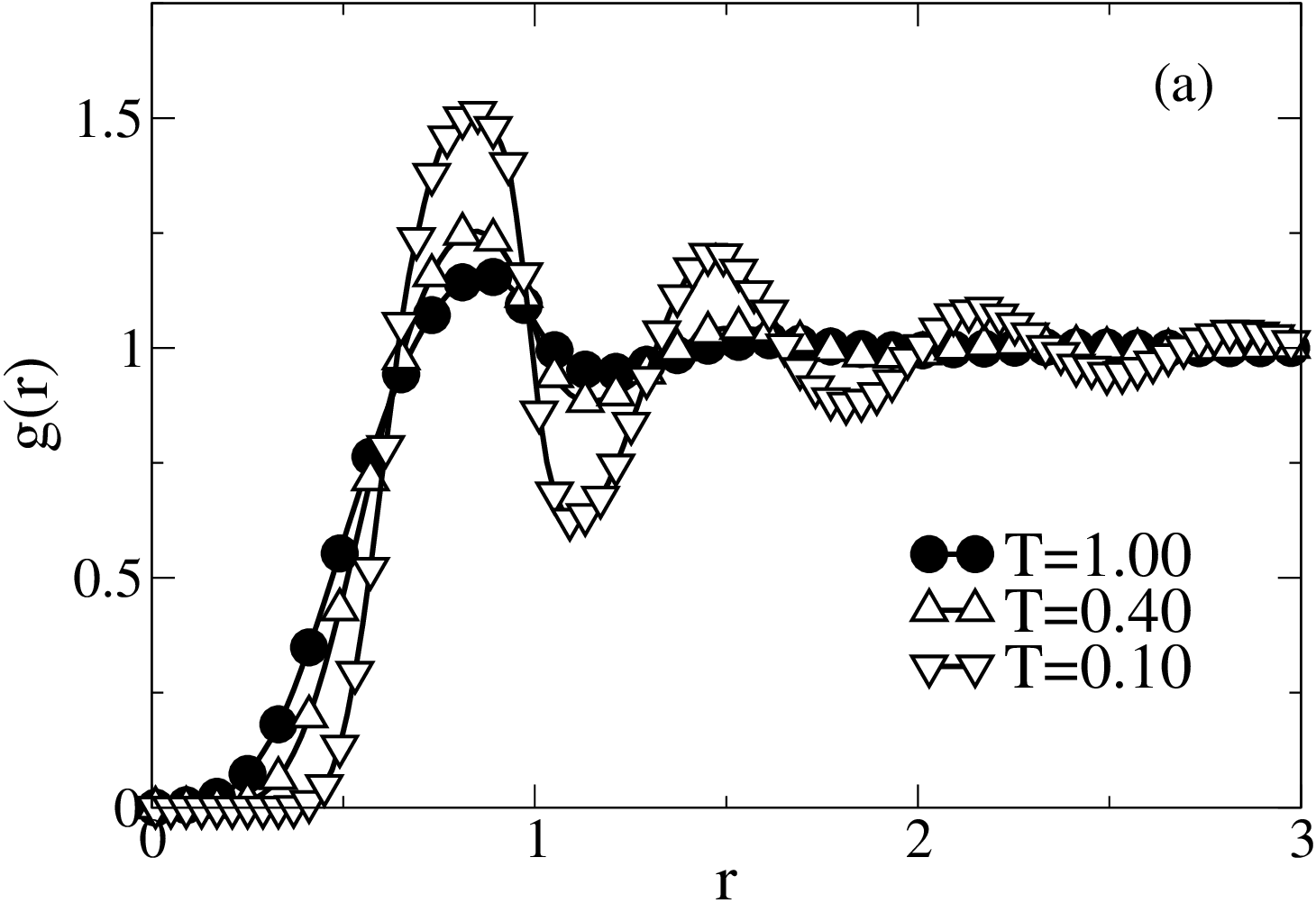}
  \includegraphics[scale=.25]{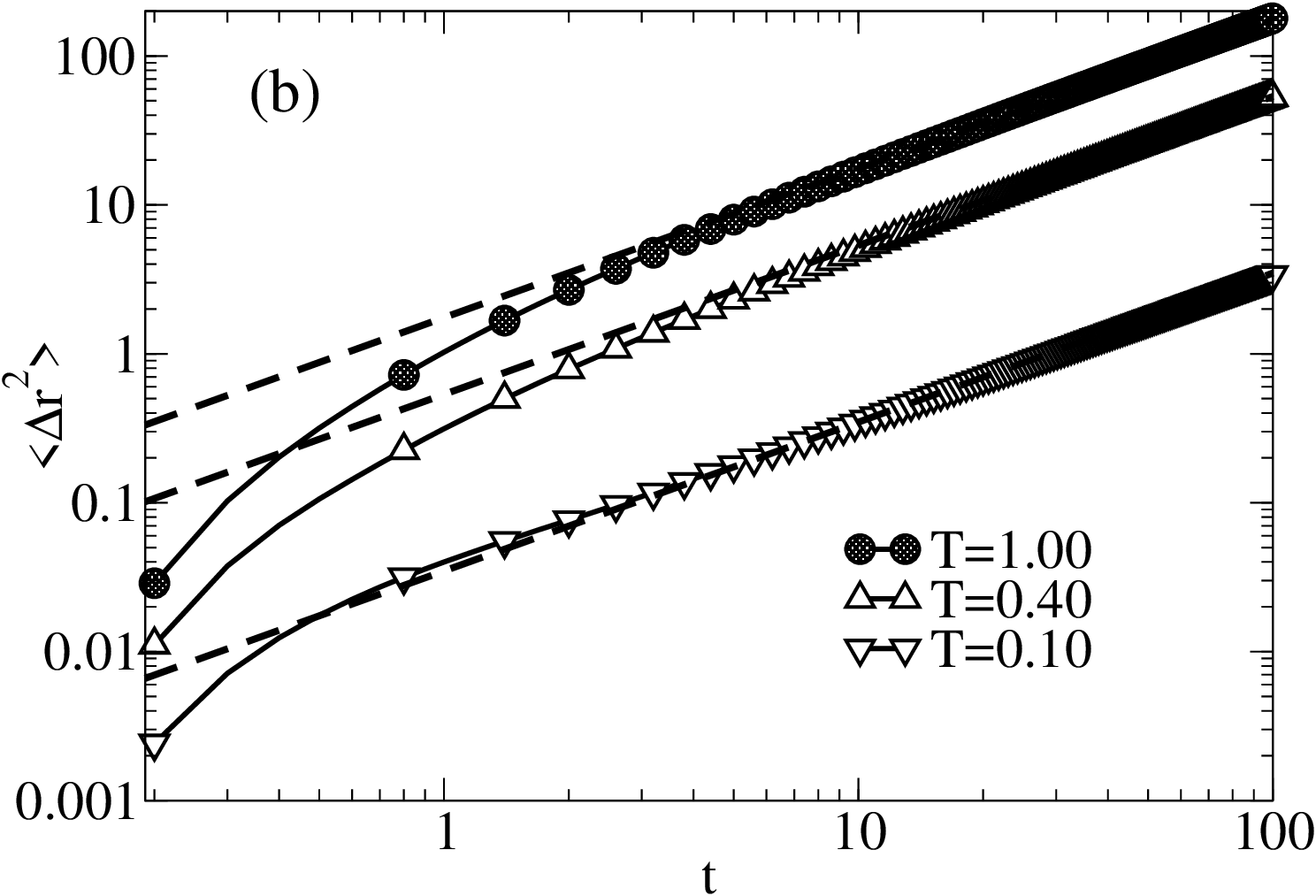}
  \caption{
    \label{fig:radial}
    (a) Radial distribution function for the DPD model at $T=1.00$,
    0.40, and 0.10. (b) Corresponding mean-square displacements
    (symbols). The dashed lines are
    $\langle \Delta r(t)^2\rangle = 6 D_s t$, where the self-diffusion
    coefficient $D_s$ (calcualted from the velocity autocorrelation
    function) is found in Table \ref{table}.  }
\end{figure}

It is informative to study the fluid structure for the different state
points investigated. Figure \ref{fig:radial}(a) plots the radial
distribution functions for three state points, namely, $T=1.00$, 0.40,
and 0.10; recall the density is always $\average{\rho}=3.0$. The
structure can be compared to the corresponding transport properties in
Table \ref{table}. First, one sees that the Schmidt number
Sc$=\nu_0/D_s$ is around 1 for $T>0.6$ and that the viscosity
decreases for decreasing temperature in the range
$0.8 \leq T \leq 1.0$, which is the well-known gas-like behavior
\cite{mcquarrie:book:1976} and also reported by Boromand et
al. \cite{boromand:cpc:2015}. In agreement with this, the radial
distribution function shows very little fluid structure in this
temperature region.

\begin{table}
  \begin{tabular}{lccccccccccc}
    \hline \hline
    $T$ & 1.00 && 0.80 && 0.60 && 0.40 && 0.20 && 0.10 \\     
    \hline
    $\eta_0^1$ & 0.715 $\pm$ 0.006 && 0.661 $\pm$ 0.008 && 0.673
    $\pm$ 0.004 && 0.778 $\pm$ 0.004 && 1.425 $\pm$ 0.008 && 4.13
    $\pm$ 0.03 \\
    $\eta_0^2$ & 0.859 $\pm$ 0.009 && 0.82 $\pm$ 0.01 && 0.848
    $\pm$ 0.005 && 1.00 $\pm$ 0.01 && 1.80 $\pm$ 0.02 && 4.83 $\pm$
    0.07 \\
    $D_s$ & 0.300 && 0.230 && 0.159 && 0.089 && 0.028 && 0.006 \\
    Sc & 1/1 && 1/1 && 1/2 && 3/4 && 17/21 &&
    229/268 \\
    $\tau_M$ & 0.075 $\pm$ 0.003 && 0.075 $\pm$ 0.002 && 0.082 $\pm$
                                                         0.003 && 0.105 $\pm$
                                                                  0.004
                                       && 0.20 $\pm$ 0.01 && 0.33 $\pm$ 0.01 \\
    $G_\infty$ & 9.5 && 8.8 && 8.2 && 7.4 && 7.1 && 12.5 \\
    \hline\hline
  \end{tabular}
  \caption{\label{table} Table of the viscosities, $\eta_0^1$ and
    $\eta_0^2$, the self-diffusivity, $D_s$, the Schmidt number,
    Sc, the Maxwell relaxation time, $\tau_M$, and instantaneous shear
    modulus, $G_\infty$. The two values for the Schmidt number are for
    $\eta_0^1/(\rho D_s)$ and $\eta_0^2/(\rho D_s)$. The
    uncertainties associated with the viscosities are the standard
    deviation of the mean calculated from five independent
    simulations. There are no statistical uncertainty on the digits
    for $D_s$, Sc. $G_\infty$ is calculated from the sample averages
    of $\eta_0^1$ and $\tau_M$ with one significant decimal.}
\end{table}
At the lowest temperature $T=0.10$, there is a clear fluid structure
and the Schmidt number is of order 10$^2$. There are no indications
that the system is crystaline for this temperature; for example, the
mean square displacement does not feature any long time plateau,
indicating no caging of the particles, and a fluidic diffusive
behavior is observed after a short time, see Fig. \ref{fig:radial}
(b). For reference, the Lennard-Jones liquid state point is
characterized by Sc $\approx$ 50. It is interesting that for $T=0.40$
a clear fluid structure is also absent in agreement with a Schmidt
number of unity and a viscosity of $\eta_0^1 = 0.70 \pm 0.01$ and
$\eta_0^2 = 0.90 \pm 0.01$ close to that of $T=1.0$.

\begin{figure}
  \includegraphics[scale=.25]{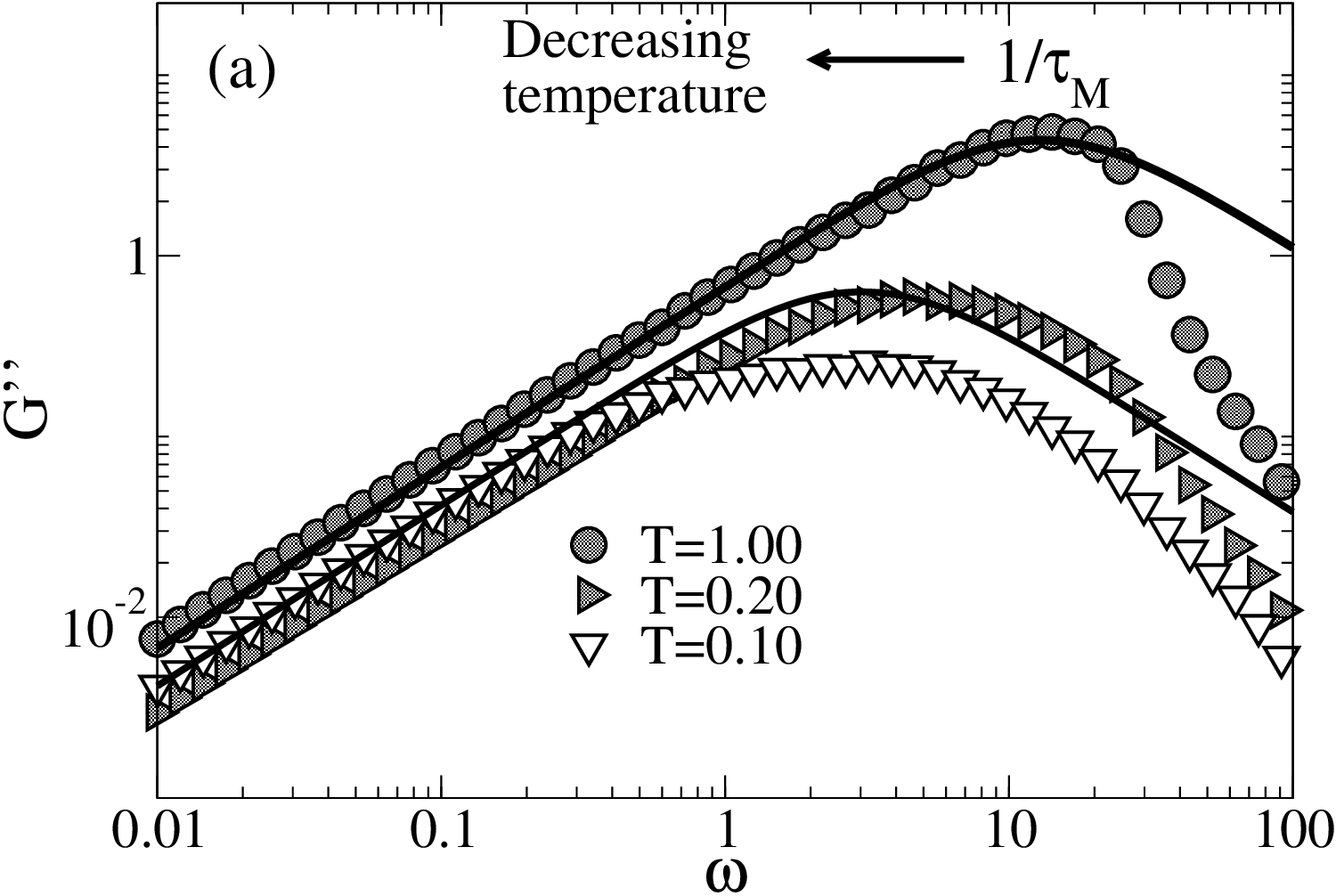}
  \includegraphics[scale=.25]{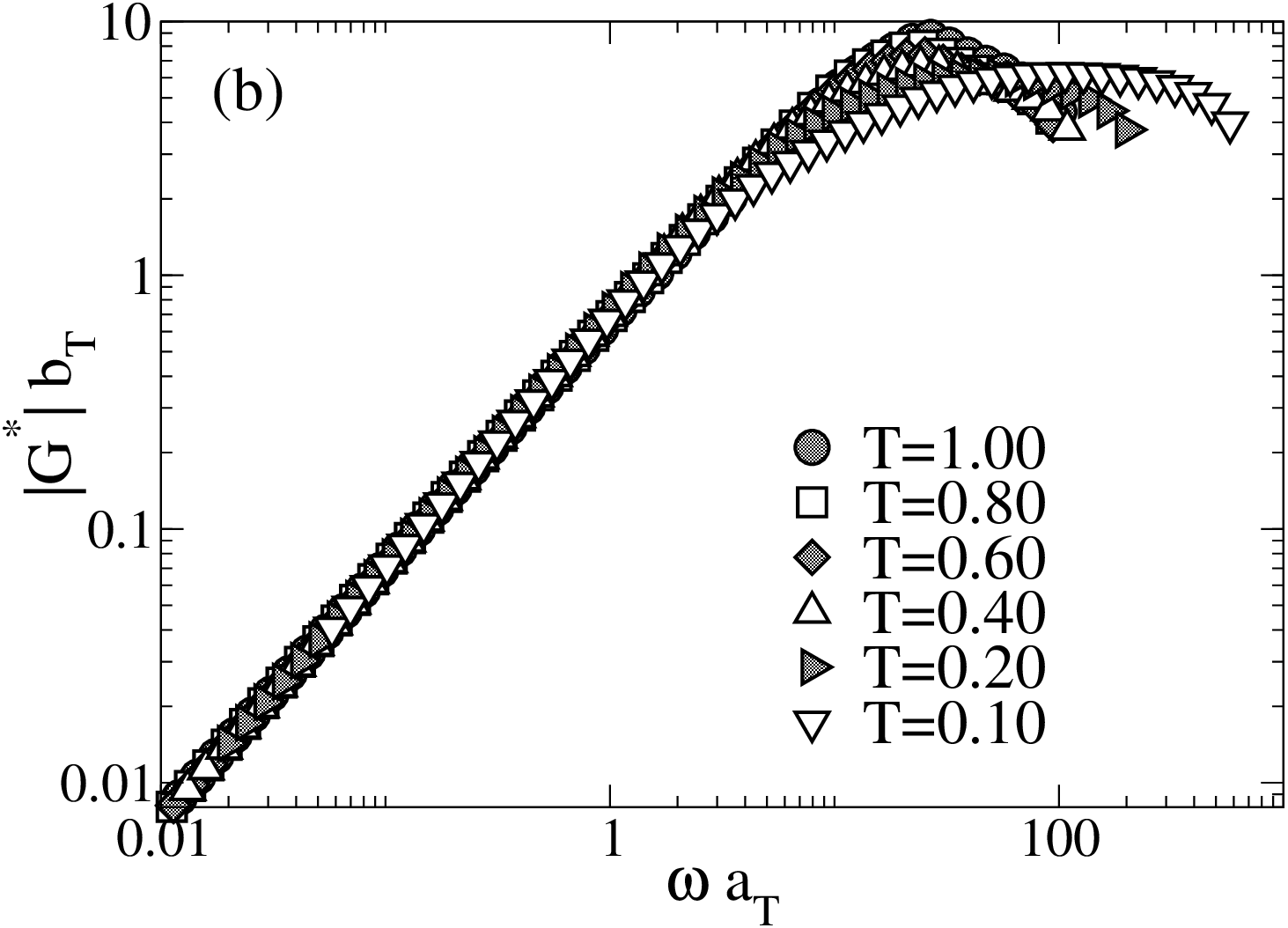}
  \caption{
    \label{fig:moduli}
    (a) The loss modulus as a function of frequency for $T=1.00, 0.20$
    and 0.10. Symbols are transformed simulation results using
    $G^*=i\omega \eta^*$, where $\eta^*$ is defined in
    Eq. (\ref{eq:etaomega}). Lines are fits to the Maxwell model, Eq.\
    (\ref{eq:maxwell}), for $T$=1.00 and 0.20. The arrow indicates
    that the inverse Maxwell time ($G''$ peak frequency) decreases for
    decreasing temperature.  (b) Test of time-temperature
    superposition using the magnitude of the shear modulus. Shift
    factors are defined as $a_T=\eta_0(T)/\eta_0(T_{\text{ref}})$ and
    $b_T=T_{\text{ref}}/T$, where $T_{\text{ref}}=1.00$.  }
\end{figure}

To study the mechanical properties further we evaluate the shear
modulus $G^* = G' + 'iG'' = i\omega \eta^*$; the loss modulus is
plotted in Fig.\ \ref{fig:moduli} for $T=1.00, 0.20$ and 0.10. Data
are compared to a single-element Maxwell model
\begin{equation}
  \label{eq:maxwell}
  G^*(\omega) = \frac{i\omega G_0}{i\omega + \tau_M^{-1}} \, ,
\end{equation}
where the Maxwell relaxation time, $\tau_M$, is found from the peak
frequency in the data and using amplitude $G_0$ as fitting
parameter. The instantaneous shear modulus (infinite-frequency complex
shear modulus), $G_\infty$, can then be found from the relation
$\eta_0/\tau_M=G_\infty$. Both $\tau_M$ and $G_\infty$ are listed in
Table \ref{table}.
From Fig.\ \ref{fig:moduli} (a) it is seen that for $T=1.00$ and
$\omega < 20$ the DPD model is Maxwellian, or equivalently, that the
shear relaxation follows a simple exponential decay for
$t>\pi/10$. For $T=0.10$ the single-element Maxwell model breaks down
at around $\omega = 0.4$. As the temperature decreases $\tau_M$
increases, thus, the shear relaxation slows down as expected. We also
test for time-temperature superposition (TTS) in Fig.\
\ref{fig:moduli}(b). Here the frequency is scaled by a factor
$a_T=\eta_0(T)/\eta_0(T_{\text{ref}})$ and the magnitude of $G^*$ by
$b_T=T_{\text{ref}}/T$ \cite{ferry:book:1980}, where the reference
temperature is $T_{\text{ref}}=1.00$. TTS applies for sufficiently low
frequencies, but fails around $\omega \approx 1/\tau_M$.
The shift factor $a_T$ increases by a factor of $\sim 6$ as temperature
decreases by an order of magnitude.

\begin{figure}
  \includegraphics[scale=.25]{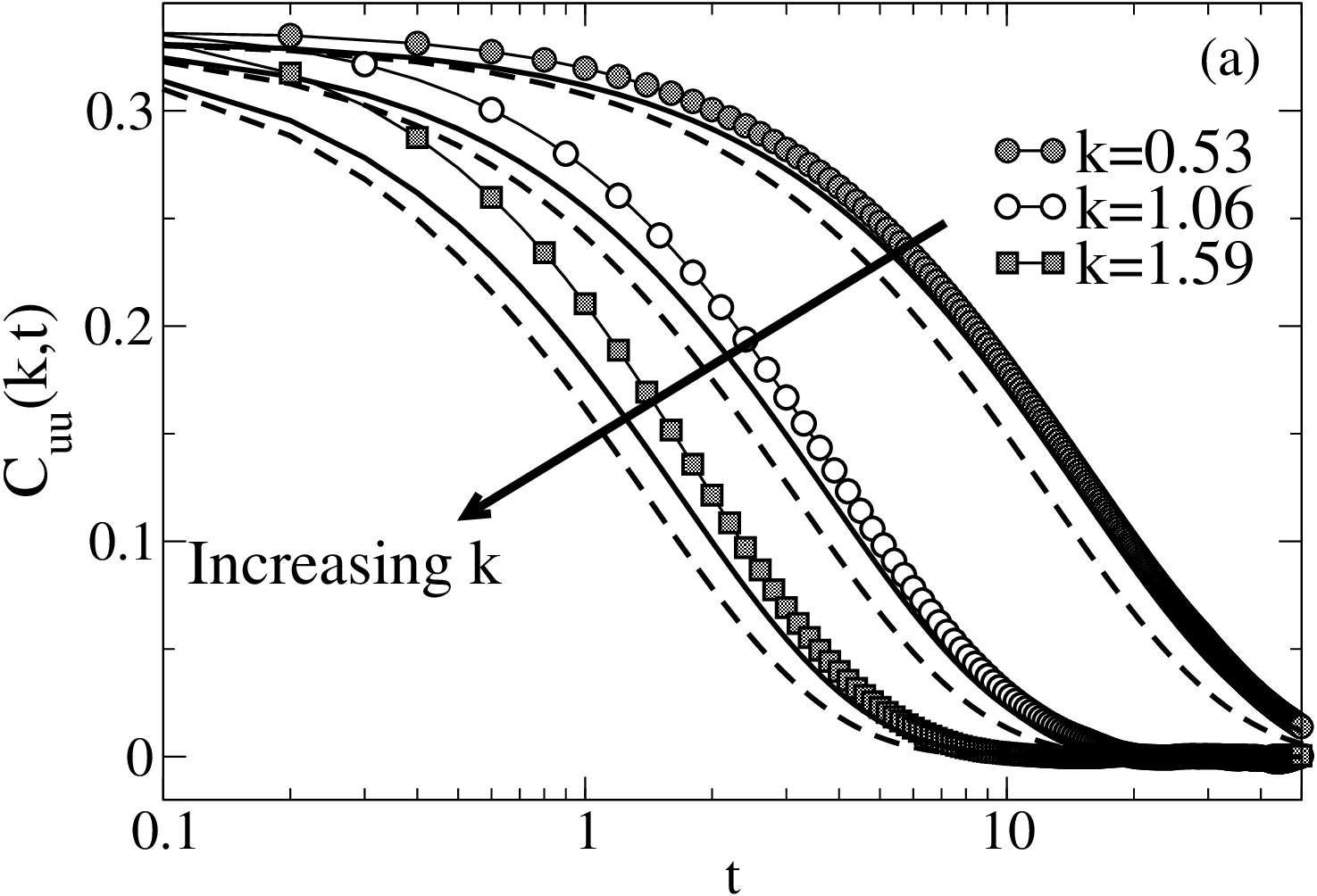}
  \includegraphics[scale=.25]{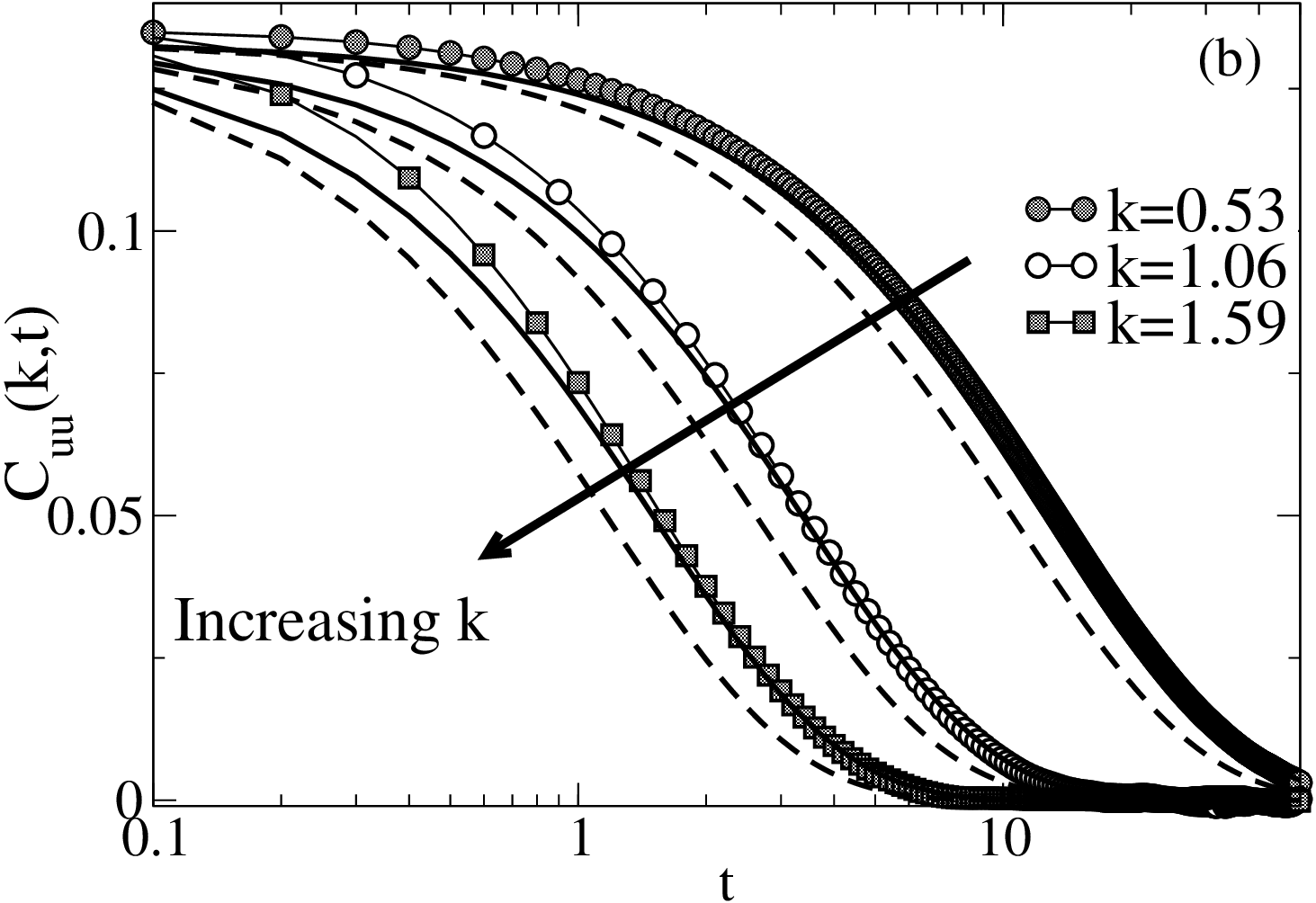}
  \caption{
    \label{fig:hydro_atoms}
    \label{fig:hydro_atomsLong}
    (a) The transverse velocity autocorrelation function,
    $C_{uu}^\perp$, in the wavevector interval $0.53 \leq k
    \leq 1.59$ for $T=1.00$. Symbols connected with lines are
    simulation results, and lines show predictions from
    Eq.\ (\ref{eq:trans:1}) using $\eta_0^1=0.715$ (full
    line) and $\eta_0^2=0.859$ (dashed line). The statistical
    uncertainty on the data are of the size of the symbols. (b) Same
    as a, but for $T=0.40$.
  }
\end{figure}

Next, we turn to the non-zero wavevector regime. We plot in
Fig.\ \ref{fig:hydro_atoms}\ (a) and (b) the transverse velocity
autocorrelation for different wavevectors at temperatures $T=1.00$ and
0.40.
It is clearly seen that the hydrodynamical theory,
Eq. (\ref{eq:trans:1}), predicts the transverse relaxation dynamics
very well in the low vector regime using the Irving-Kirkwood
definition of the pressure tensor. Applying the Jung-Schmid definition
gives too fast a relaxation, which indicates that this particular
dynamical mode is not dependent on the random and dissipative shear
forces. More quantitatively: the theory predicts the half-life
as $t_{1/2}= \ln(2)/(\nu_0 k^2)$, i.e. for $T=0.4$ we have
$t_{1/2}=9.5$ using $\eta_0^1=0.78$ and $t_{1/2}=7.4$ using
$\eta_0^2=1.00$. This can be compared to the simulation result
$t_{1/2}=9.5$. For very short times the theory fails to predict the
relaxation; this is to be expected as the viscosity is in general both
frequency and wavevector dependent, hence, for sufficiently short
times the time dependence of the viscosity is important.

\begin{figure}
  \includegraphics[scale=.25]{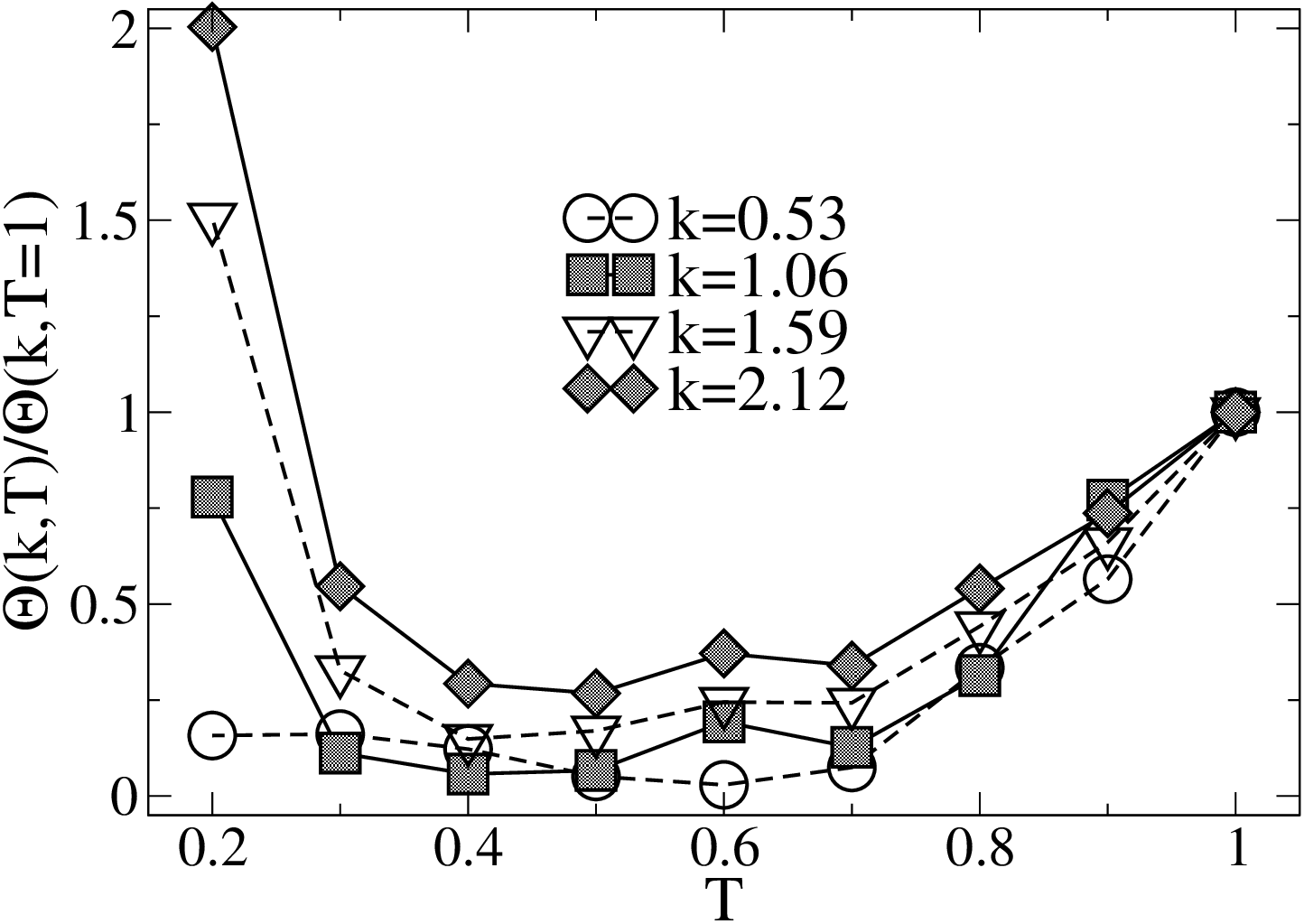}
  \caption{
    \label{fig:reldeviation}Normalized mean sqaure deviation $\Theta$
    as a function of temperature and for different
    wavevectors. Lines serve as a guide to the eye. 
  }
\end{figure}

Interestingly, the agreement is less satisfactory for $T$=1.0; here
the Irving-Kirkwood definition yields $t_{1/2}$ = 10.4 versus
the simulation result $t_{1/2}=11.2$. In Fig. \ref{fig:reldeviation}
we plot the mean square deviation
\begin{equation}
  \Theta(k, T) = \frac{1}{N_s}\sum_i
  \left(
  \frac{\average{\rho}}{k_BT} C_{uu}^\perp(k, t_i) - \frac{C_{uu,
      k,t_i}^\perp}{C_{uu, k,0}^\perp} 
  \right)^2
\end{equation}
where $C_{uu}^\perp(k, t_i)$ is the predictions from the theory, and
$C_{uu, k,t_i}^\perp$ simulation data. To avoid this parameter being
affected by the large noise-to-signal ratio at very long times, we
only sum over the $N_s$ times with data points 
$C_{uu}^\perp(k,t)/C_{uu}^\perp(k,0) \geq 0.1$.
Clearly, the mininum deviation is found within the temperature region
$0.3 \leq T \leq 0.7$. For higher temperatures the agreement is not as
satisfactory; here we approach the kinetic regime as defined by
Espa\~{n}ol and Serrano \cite{espanol:pre:1999}, that is, low friction
and overlap parameters mentioned above. For low temperatures one
observes a quite large deviation, especially pronounced for larger
wavevectors. This, we argue, is due to the large characteristic
frequency, $\omega = \nu_0 k^2$, which is outside the classical
hydrodynamic regime. For $T=0.1$ this hydrodynamic regime is never
reached because of the limitations on the wavevector
$k_{\mathrm{min}}=2\pi/L_{\mathrm{box}}$.

\begin{figure}
  \includegraphics[scale=.25]{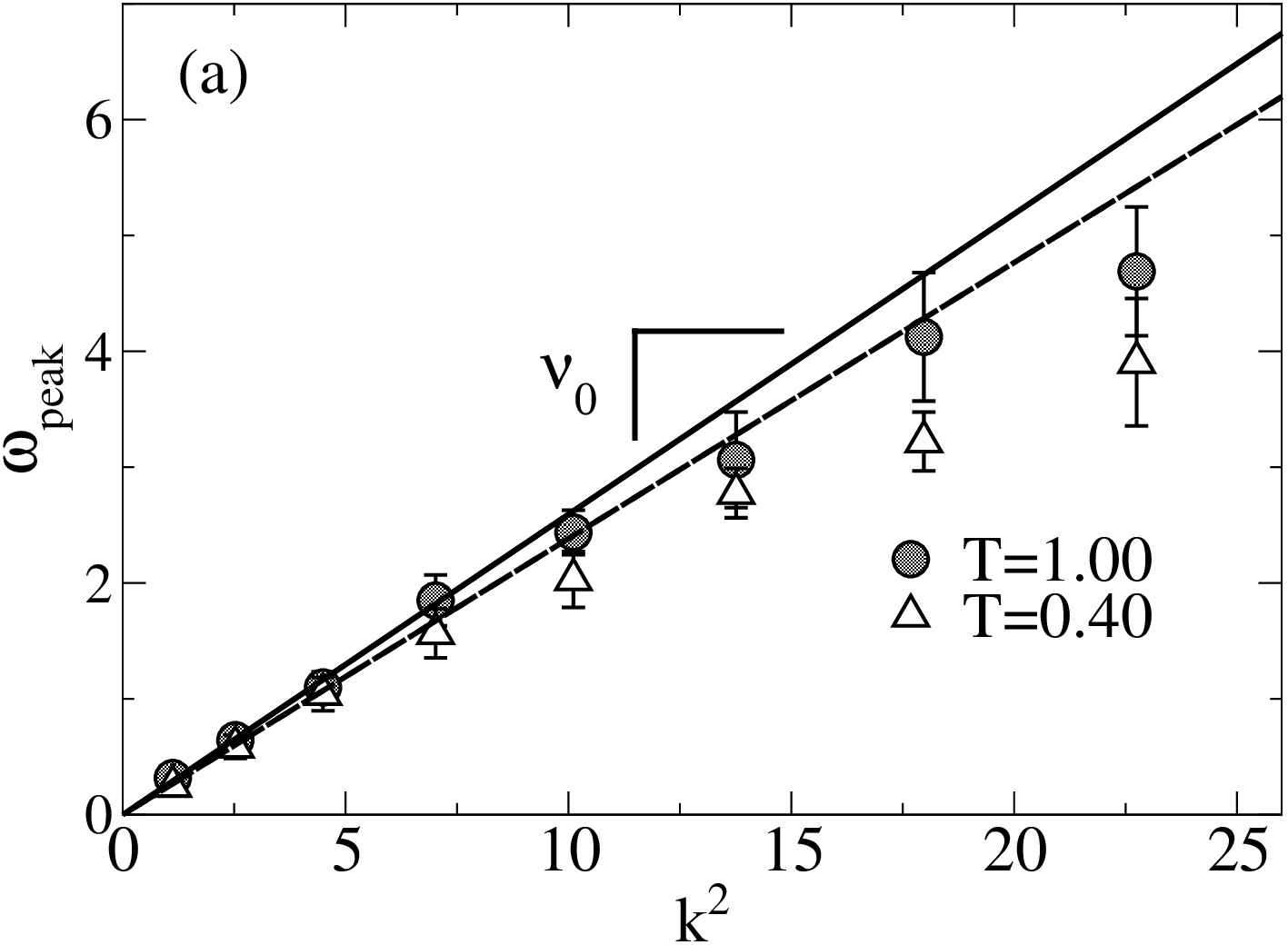}
  \includegraphics[scale=.25]{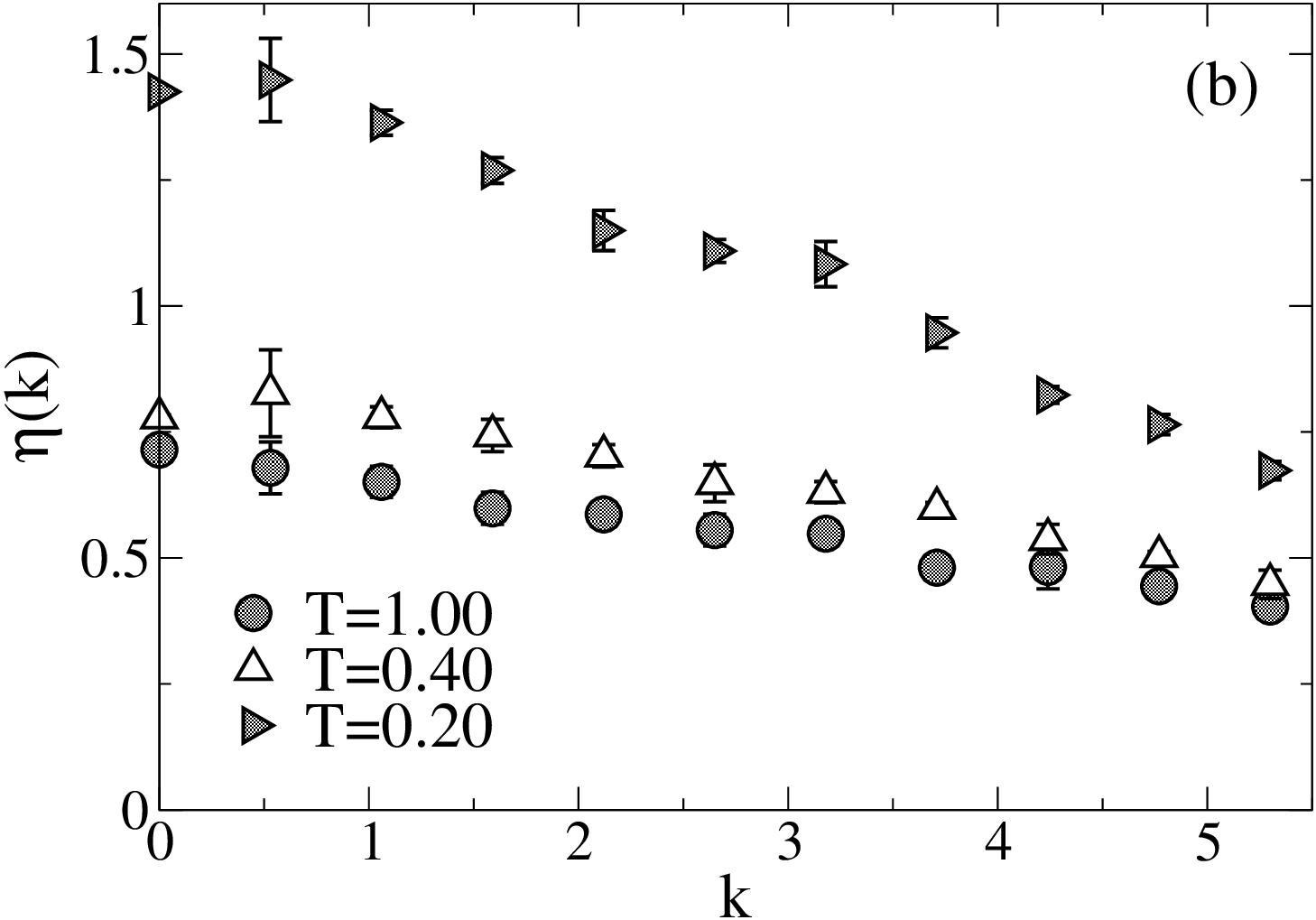}
  \caption{
    \label{fig:peak-kernel}
    (a) Dispersion relations for $\omega_{\mathrm{peak}}$ for
    temperatures $T=1.00$ and $T=0.40$. The lines are hydrodynamic
    predictions; the viscosity is given by the slope. (b) Viscosity
    kernels for $T=1.00, 0.40$ and 0.20.  }
\end{figure}

Fourier-Laplace transformation of Eq.\ (\ref{eq:trans:1}) leads to 
\begin{equation}
  \label{eq:sss}
  \widehat{C}_{uu}^{\perp}(\vec{k},\omega)=
  \frac{k_BT}{\average{\rho}}\int_0^\infty e^{-\nu_0 k^2 t}
  e^{-i\omega t} \hardd t = \frac{k_BT}{\rho}\frac{1}{\nu_0 k^2 + i\omega} \, ,
\end{equation}
which gives a peak in the imaginary part of the spectrum at
$\omega_{\text{peak}}=\nu_0k^2$. This peak frequency found from the
simulations is plotted in Fig.\ \ref{fig:peak-kernel} (a) for $T=1.00$
and $T=0.40$ together with the hydrodynamic predictions. For low
wavevectors, the peak frequency follows the predictions:
$\omega_{\text{peak}}$ is proportional to $k^2$ and the relaxation is
governed by the diffusion of momentum. The prediction fails for larger
wavevectors; at lower temperature the deviation is significant for
relatively lower wavevectors compared to high temperature. Again, we
attribute this to the large characteristic frequency at low
temperature and large wavevector.
The frequency and wavevector dependent shear viscosity can be defined
by re-arranging Eq.\ (\ref{eq:sss}),
\begin{equation}
  \widehat{\eta}(k,\omega)=\frac{k_BT-i\omega \rho 
    \widehat{C}_{uu}^{\perp}(\vec{k},\omega)}{ k^2
    \widehat{C}_{uu}^{\perp}(\vec{k},\omega)} \, .
\end{equation}
This result can also be derived from first principles by including the
position and time dependence of the transport coefficient in
Eqs. (\ref{eq:const}). In the zero frequency limit we have the
viscosity kernel
$\w{\eta}(k) = k_BT/k^2\widehat{C}_{uu}^{\perp}(\vec{k},0)$.  Figure
\ref{fig:peak-kernel} (b) shows this viscosity kernel at zero
frequency for $T=1.00, 0.40$ and $T=0.20$. The zero wavevector
viscosity is also indicated using $\eta_0^1$ from Table
\ref{table}. It is interesting to see that for $k$ less than unity,
the wavevector-dependent viscosity reaches $\eta_0$, i.e., 
the local Newtonian law of viscosity holds for $k<1.0$. This is observed (in
appropriate reduced units) for many different fluids
\cite{hansen:langmuir:2015}. We also note that Ripoll et
al. \cite{ripoll:jcp:2001} studied the kernel for zero conservative
force. 

Rather than approaching the deviation between theory and simulation
through wavevector dependent transport coefficients, one can
generalize the stochastic forcing and assume ${\fluc{\vec{J}}}^m$,
${\fluc\tensor{P}}$, and ${\fluc{\tensor{J}}}^e$ to be correlated with
hydrodynamic quantities.  In this case the transverse dynamics are
governed by the equation
\begin{equation}
   \frac{\partial C_{uu}^\perp}{\partial t} = -\nu_0 k^2C_{uu}^\perp +
   \varepsilon(\vec{k}, t)
\end{equation}
where
\begin{equation}
   \varepsilon(\vec{k}, t) = -
   \frac{ik}{\average{\rho}V} \langle \w{\fluc{P}}_{yx}(\vec{k},t)
   \w{\fluc{u}}_x(-\vec{k},0) \rangle \, \neq 0 . 
\end{equation}
Applying a Fourier-Laplace transform gives the correlation between forcing and
the transverse velocity in terms of wavevector and frequency as
\begin{equation}
  \widehat{\varepsilon}(\vec{k}, \omega) = (i\omega + \nu_0k^2)
  \widehat{C}_{uu}^{\perp}(\vec{k},\omega) - C_{uu}^\perp(\vec{k},0)
  \, . 
\end{equation}
Because the theoretical predictions are relatively large for higher
temperatures the contribution from $\varepsilon$ is larger for all
wavevectors compared to the intermediate temperatures. 

\begin{figure}
  \includegraphics[scale=0.3]{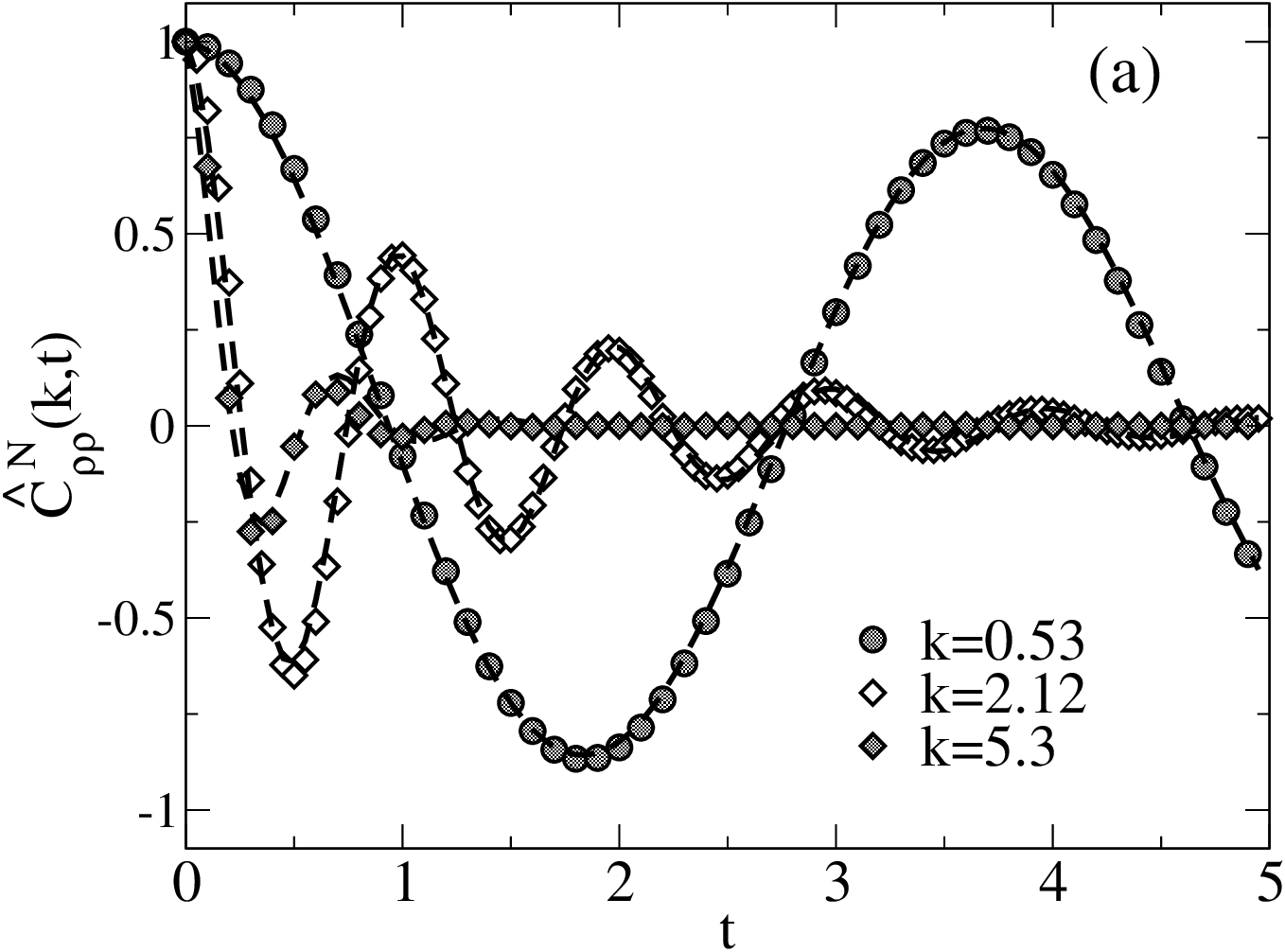}
  \includegraphics[scale=0.3]{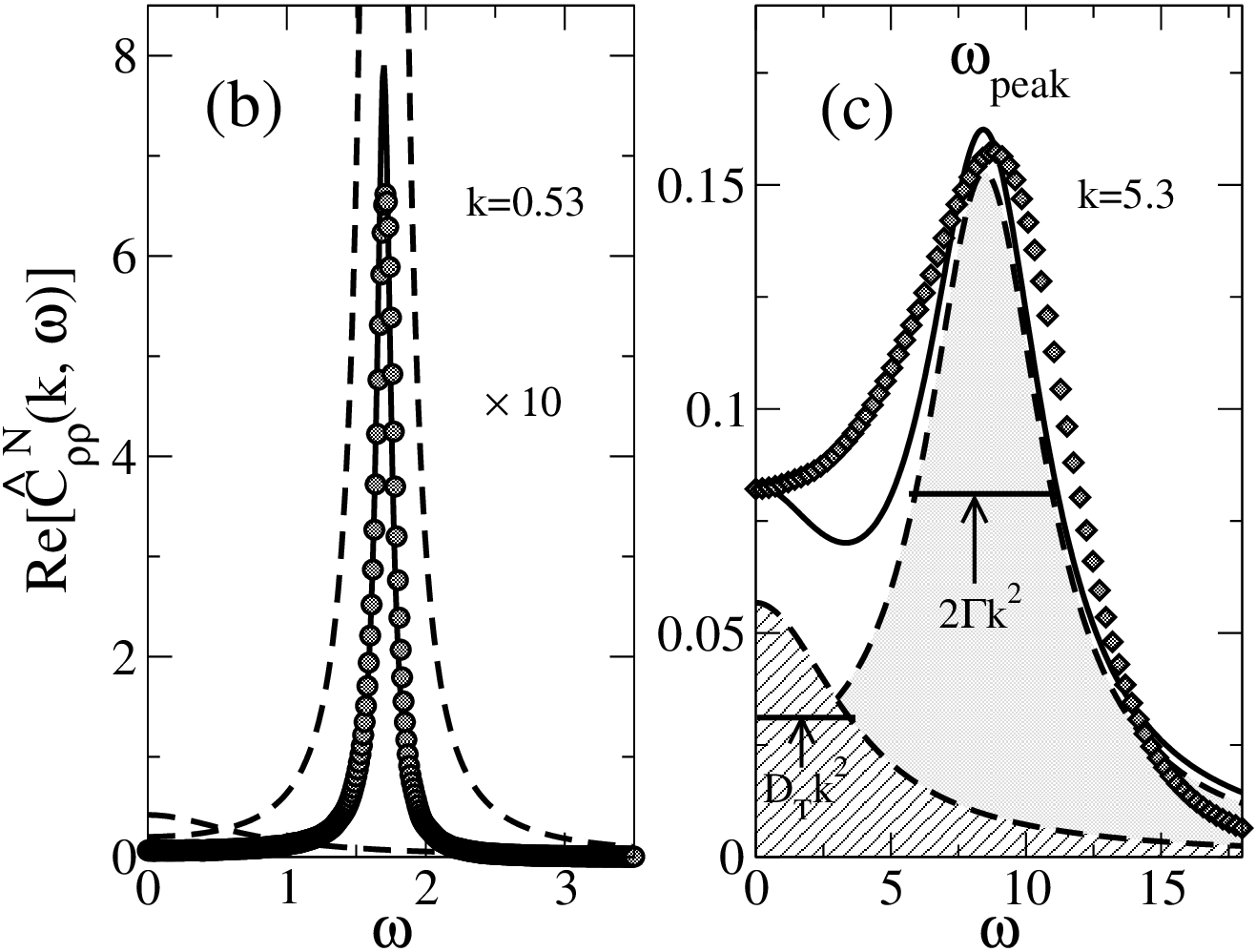}
  \caption{
    \label{fig:lacfT1}
    (a): The density autocorrelation function for
    wavevectors $k=0.53, 2.12$ and 5.30 at $T=1.00$. Symbols 
    are the simulation results and dashed lines the best fit of
    Eq. (\ref{eq:longnorm}) to data. (b) and (c): Spectra of the
    density autocorrelation function for $k=0.53$ and $k=5.3$,
    respectively, at $T=1.00$. Symbols are Fourier-Laplace transformed data
    points. The dashed lines are the Rayleigh and Brillouin terms,
    Eq. (\ref{eq:longrromega}); in (b) these contribution are
    multiplied by a factor 10 for clarity. The shaded areas, (c),
    indicate the Rayleigh and Brillouin integral regions.  }
\end{figure}

We now turn to the longitudinal relaxation dynamics and focus first on
the density autocorrelation function, $C_{\rho \rho}$. It is worth
noting that this is related to the coherent intermediate scattering
function, $F(\vec{k},t)$, by
$C_{\rho\rho}(\vec{k},t)=\average{\rho}F(\vec{k},t)$. The density
autocorrelation function is a real-valued function and, hence, it
relaxes according to Eq.\ (\ref{eq:longnorm}); it is plotted in
Fig.\ \ref{fig:lacfT1} (a) for $T=1.00$ at wavevectors $k=0.53,
2.12$ and 5.30.
The dashed line is the best fit of Eq.\ (\ref{eq:longnorm}) to data
using $K_{\rho \rho}, D_T, \Gamma$ and $c_s$ as fitting
parameters. The damped oscillations predicted from hydrodynamics are
evident, indicating sound waves that are dampened by the
sound attenuation coefficient, $\Gamma$. The existence of the thermal
process is less clear. To study this in more detail we investigate the
corresponding spectra. The Fourier-Laplace transform of
Eq.\ (\ref{eq:longnorm}) is 
\begin{equation}
  \widehat{C}_{\rho\rho}^N(\vec{k},\omega)=\frac{K_{\rho\rho}}{D_T k^2 +
    i\omega} - \frac{(1-K_{\rho\rho})(i\omega - \Gamma k^2)}{(c_sk)^2 +(i\omega-
    \Gamma k^2)^2} \, .
  \label{eq:longrromega}
\end{equation}
Again, note the two different contributions to the relaxation. The
real part of $\widehat{C}_{\rho\rho}^N$ is symmetric about $\omega=0$
and we therefore only discuss the behaviour for $\omega \geq 0$. The
first term relates to the thermal process and gives rise to the
Rayleigh peak at $\omega=0$; this process is only present at low
frequencies and the half-width of the Rayleigh peak is $D_Tk^2$. The
second term has a peak at frequency $\omega_{\mathrm{peak}}=c_sk$; the
maximum is identified as the Brillouin peak and has half width
$2\Gamma k^2$. Inspired by Hansen and McDonald
\cite{hansen:book:2006}, the peaks and their widths are illustrated in
Fig. \ref{fig:lacfT1} (c). Figures \ref{fig:lacfT1} (b) and (c) show
the real part of the spectrum of the density autocorrelation function
for wavevectors $k=0.53$ and $k=5.30$, the highest wavevector
studied. Using the fitted values obtained in Fig. \ref{fig:lacfT1}
(a), we plot the predicted spectra together with the transformed
data. The agreement is not perfect as the local minimum predicted by
the theory (at $\omega \approx 4$ for $k=5.3$) is not found in the
spectrum of the data. Fitting to Eq. (\ref{eq:longrromega}) did not
improve this. For $T=1.0$ and relatively small wavevectors,
Fig. \ref{fig:lacfT1} (b), the thermal process is almost completely
absent and the relaxation is athermal. However, for large wavevector,
Fig. \ref{fig:lacfT1} (c), the process is indeed observed in the
spectrum.

\begin{figure}
  \includegraphics[scale=0.35]{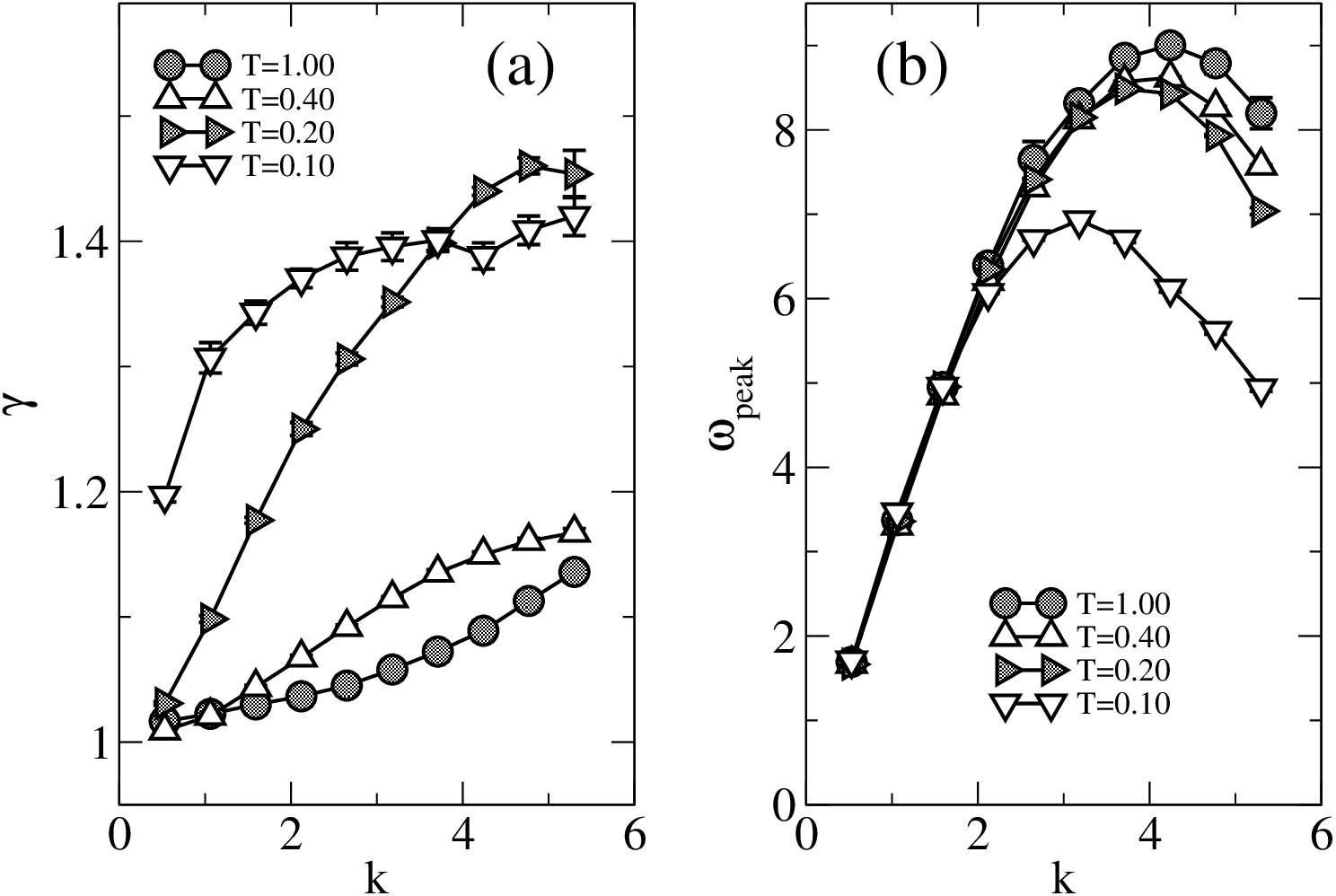}
  \caption{
    \label{fig:lp}
    (a) Dispersion relation for $\gamma = I_R/2I_B + 1$ for different
    temperatures. (b) The corresponding dispersion relation for the 
    peak frequency $\omega_{\mathrm{peak}}$; the hydrodynamic
    prediction is $\omega_{\mathrm{peak}}=c_s k_y$.
  }
\end{figure}

The ratio of the two processes is quantified from the
Landau-Placzek ratio \cite{cummins:jcp:1966}, that is, the ratio
between the Rayleigh and Brillouin integral regions, or intensities, 
$I_R/2I_B=\gamma-1$, where $\gamma$ itself is the ratio between the
heat capacities at constant pressure and volume, $\gamma = C_P/C_v$. The
integral regions are also illustrated in Fig. \ref{fig:lacfT1} (c). In
Fig. \ref{fig:lp} (a) we plot the dispersion relation for $\gamma$ for
different temperatures. 
It is clear that the thermal process intensity increases as we
decrease temperature and wavelength. For reference, the Lennard-Jones
liquid features $ 1.6 \leq \gamma \leq 2.6$ for $ 0.46 \leq k \leq 5.9
$, see also Bryk et al. \cite{bryk:jcp:2010}. In this region, the
Lennard-Jones system also shows a clear de Gennes narrowing
\cite{hansen:book:2006}; we have not observed this narrowing for the
wavevectors and temperatures studied here. From Fig. \ref{fig:lacfT1} (c)
it is seen that the frequencies of two processes overlap indicating
that the processes are coupled; this coupling is only present on
relatively small length scales, that is, for typical simulation setups
these two processes are decoupled and, furthermore, the thermal
process only accounts for a small fraction of the hydrodynamic
relaxation. However, for $T=0.10$, the coupling is relatively large
even on longer length scales and may affect the response considerably.

The dispersion relation for the Brilluion peak frequency,
$\omega_{\text{peak}}$, is plotted in Fig. \ref{fig:lp} (b); it is seen
that the oscillatory frequencies are roughly the same for the
different temperatures at sufficiently small wavevector, which means
that the speed of sound is to a good approximation independent of
temperature on these length scales. For larger wavevectors the
discrepancy between $T=0.10$ and $T>0.10$ is pronounced; the
underlying mechanical reason for this is not well known, but likely
due to the different local liquid structure on these small scales, see
for example Ref. \onlinecite{trachenko:rpp:2016}, but also the
coupling of the longitudinal processes can be important. It is
worth noting that the maxima seen in Fig. \ref{fig:lp} (b) is also
observed in the Lennard-Jones liquid. From the simulation data we
cannot conclude if the DPD model features positive or negative
dispersion \cite{bryk:jcp:2010,trachenko:rpp:2016}.

\begin{figure}
  \includegraphics[scale=.25]{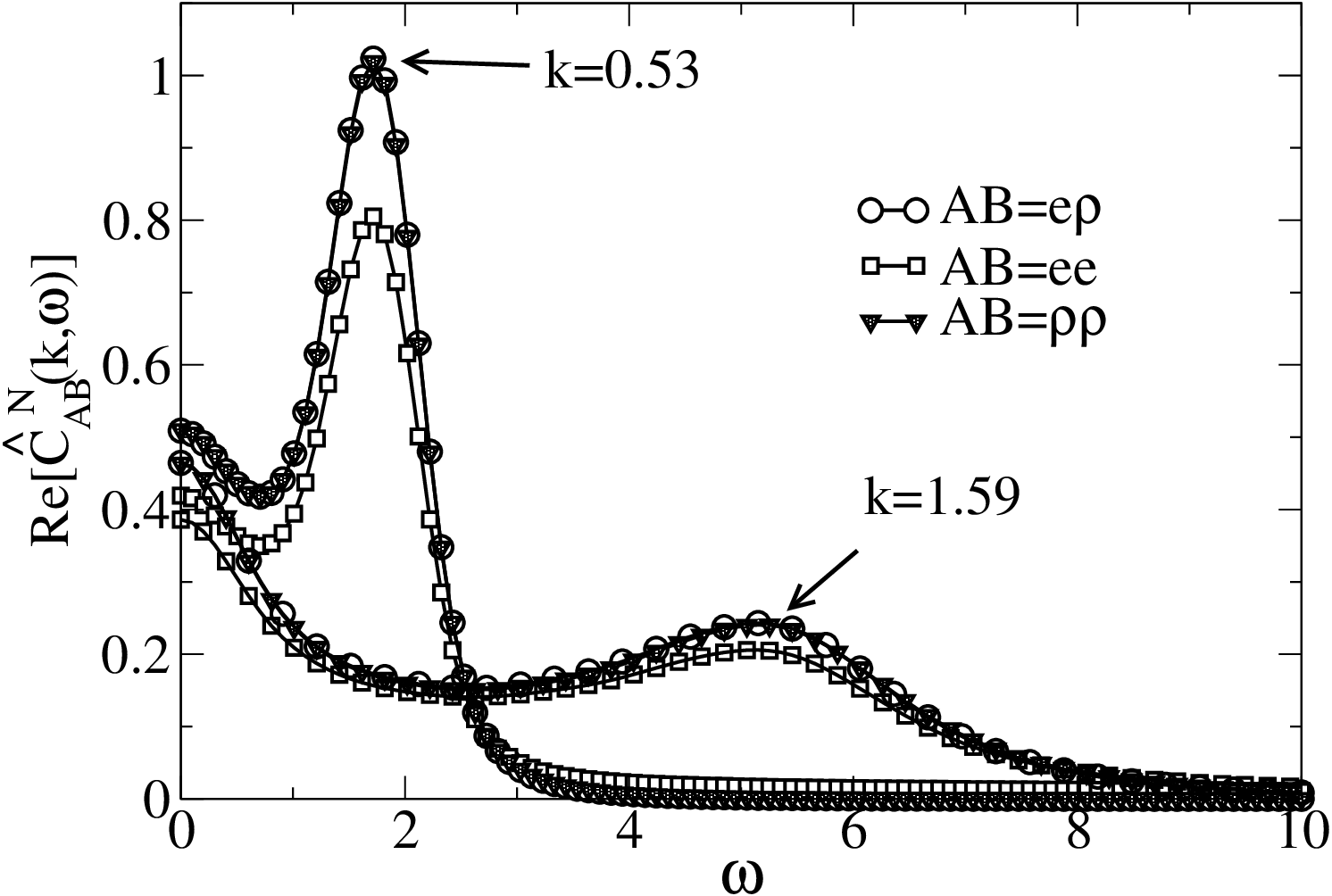}
  \caption{
    \label{fig:atoms:long}
    (a) $\widehat{C}_{\rho\rho}^N, \widehat{C}_{e\rho}^N$, and
    $\widehat{C}_{ee}^N$ for wavevectors $k=0.53$ and $k=1.59$.}
\end{figure}

We conclude our investigation of the collective properties by plotting
in Fig.  \ref{fig:atoms:long} the density-density, density-energy, and
energy-energy correlation functions at $T=0.10$ for two different
wavevectors.
It is seen that these three different correlation functions have the
same characteristics as discussed above, in agreement with the
hydrodynamic predictions. That is, the standard DPD system includes
the cross coupling between the longitudinal quantities hydrodynamically, at
least qualitatively.

\section{Summary and Conclusions}
In this paper the equilibrium relaxations of the standard dissipative
particle dynamics model propopsed by Groot and Warren
\cite{groot:jcp:1997} were investigated. First, the well-known results
that the structure and dynamics at high temperatures ($T \geq 0.8$)
resemble those of a gas were recaptured; this region in phase space is
accordingly denoted the kinetic regime \cite{espanol:pre:1999}. At
lower temperatures the viscosity increases with decreasing temperature
and the Schmidt number approaches that of the model Lennard-Jones
liquid. The DPD model features a single element Maxwellian shear modulus
relaxation behavior for sufficiently small frequencies that depend on
the temperature; the lower the temperature the smaller the frequencies that are
required to observe Maxwellian behavior. Also, the time-temperature
superposition principle is applicable in the low frequency regime.

For nonzero wavevectors, the hydrodynamic prediction for the
transverse velocity autocorrelation function is tested using the
Jung-Schmid and the Irving-Kirkwood definitions of the viscosity; the
former includes the random and dissipative shear force contributions
whereas the latter only includes the conservative and kinetic
contributions. Using the Irving-Kirkwood viscosity the hydrodynamic
predictions are in excellent agreement with simulations results for
temperatures $0.4 \leq T \leq 0.7$ and $0.53 \leq k_y \leq
2.12$. Importantly, using the Jung-Schmid viscosity overestimates the
relaxation, indicating that the transverse relaxation dynamics are
independent of the dissipative and random shear forces. Also, for
higher temperatures the agreement is less satisfactory, for a given
wavevector, in accordance with the Bocquet-Chaix criterion.

A qualitative investigation into the longitudinal dynamics was also
carried out. For the high temperature regime ( $T \geq 0.8$), the
density longitudinal spectrum at low wavevectors is characterized by a
single sharp Brillouin peak. This indicates that the longitudinal
relaxation is athermal and dominated by propagating damped density
waves. This mechanism is very different compared to a simple liquid,
in which the thermal diffusion process dominates at low wavevector. In
the low temperature range, the Rayleigh peak is more prominent; a
fingerprint of the thermal diffusion process. Dispersion relations for
the Landau-Placzek ratio shows that the thermal process intensity
increases compared to the wave propagation process as the length scale
decreases; this is true for all temperatures and wavevectors studied
and also the case for the Lennard-Jones liquid, even though the
Landau-Placzek ratio is larger here. For the supercritical fluid
Lennard-Jones model there is a small increase in the speed-of-sound
with respect to temperature \cite{bryk:jcp:2010}, however, for the DPD
model this a constant with respect to temperature for $k<2$. Finally,
the DPD model features the cross couplings predicted by the theory, at
least, qualitatively.

In conclusion, the thermal fluctuations in the standard coarse grained
DPD model by Groot and Warren \cite{groot:jcp:1997} preserves, at
least qualitatively, the underlying mechanical processes predicted by
classical hydrodynamic theory. Therefore, the model can be used to
study fluctuating hydrodynamics as stated by Espa{\'{n}}ol and
Warren \cite{espanol:jcp:2017}. However, we suggest to use low
temperature settings where $T \leq 0.7$.

\section{Acknowledgments}
Innovation Fund Denmark supported this work as a part of the ROSE
project (no. 5160-00009B).  MLG acknowledges support from the US
Fulbright Program. This work was supported by the VILLUM Investigator
grant for the Matter project (JCD). Finally, we wish to thank the
reviewers for their valuable comments. 

\bibliographystyle{unsrt}


\end{document}